\documentclass[12pt]{article}
\usepackage{amsmath, amsfonts}
\usepackage{siunitx}
\usepackage{graphicx,psfrag,epsf}
\usepackage{enumerate}
\usepackage{natbib}
\usepackage{amsthm}
\usepackage{bm}
\newtheorem{theorem}{Theorem}
\usepackage{url} 
\usepackage[colorlinks,citecolor=blue,urlcolor=blue,filecolor=blue]{hyperref}
\usepackage{makecell}

\newcommand{\blind}{1}

\begin{document}

\def\spacingset#1{\renewcommand{\baselinestretch}%
{#1}\small\normalsize} \spacingset{1}


\if1\blind
{
  \title{\bf  Bayesian Nonparametric Clustering with Feature Selection for Spatially Resolved Transcriptomics Data}
  
  \author{Bencong Zhu \\ 
    Department of Statistics, The Chinese University of Hong Kong \\
    Guanyu Hu \\
    Department of Biostatistics and Data Science and Center\\
    for Spatial Temporal Modeling for Applications in Population Sciences,\\ The University of Texas Health Science Center at Houston\\
    Yang Xie \\
    Peter O'Donnell Jr. School of Public Health,\\ The University of Texas Southwestern Medical Center\\ 
    Lin Xu \\
    Peter O'Donnell Jr. School of Public Health,\\ The University of Texas Southwestern Medical Center\\ 
    Xiaodan Fan$^{*}$ \\
    Department of Statistics, The Chinese University of Hong Kong  \\
    and \\
    Qiwei Li$^{*}$ \\
    Department of Mathematical Sciences, The University of Texas at Dallas \\
    }
  \maketitle
  \clearpage
} \fi

\if0\blind
{
  \bigskip
  \bigskip
  \bigskip
  \begin{center}
    {\LARGE\bf Bayesian Nonparametric Clustering with Feature Selection for Spatially Resolved Transcriptomics Data}
\end{center}
  \medskip
} \fi

\bigskip
\begin{abstract}
The advent of next-generation sequencing-based spatially resolved transcriptomics (SRT) techniques has reshaped genomic studies by enabling high-throughput gene expression profiling while preserving spatial and morphological context. Nevertheless, there are inherent challenges associated with these new high-dimensional spatial data, such as zero-inflation, over-dispersion, and heterogeneity. These challenges pose obstacles to effective clustering, which is a fundamental problem in SRT data analysis. Current computational approaches often rely on heuristic data preprocessing and arbitrary cluster number prespecification, leading to considerable information loss and consequently, suboptimal downstream analysis. In response to these challenges, we introduce BNPSpace, a novel Bayesian nonparametric spatial clustering framework that directly models SRT count data. BNPSpace facilitates the partitioning of the whole spatial domain, which is characterized by substantial heterogeneity, into homogeneous spatial domains with similar molecular characteristics while identifying a parsimonious set of discriminating genes among different spatial domains. Moreover, BNPSpace incorporates spatial information through a Markov random field prior model, encouraging a smooth and biologically meaningful partition pattern. We assess the performance of BNPSpace  utilizing both simulated and real SRT data. In the application to the human dorsolateral prefrontal cortex (DLPFC) 10x Visium data, comprising $4,788$ spots and $33,538$ genes with a $95\%$ sparsity, BNPSpace outperforms existing methods by identifying more coherent spatial domain patterns and achieving the best clustering performance. Furthermore, the discriminating genes identified by BNPSpace showed significant enrichment with odd ratio $=1.877$ ($p$-value = $0.00162$) in DLPFC gene sets validated by real biological experiments, underscoring its effectiveness in revealing biologically relevant insights.

\end{abstract}

\noindent%
\vfill

\newpage
\spacingset{1.45} 
\section{Introduction}
\label{sec:intro}

The advent of single-cell RNA sequencing (scRNA-seq), a next-generation sequencing (NGS) method, has revolutionized our comprehension of gene expression regulation across various cell lineages, thereby shedding light on their crucial roles in development and diseases \citep{tanay2017scaling,papalexi2018single}. Despite its transformative impact, the process of tissue dissociation in scRNA-seq results in the loss of spatial context in gene expression, which is crucial for understanding biological functions \citep[e.g., cell-cell and cell-environment interactions;][]{gruene2011laser}, biological process \citep[e.g., embryo development and tumor progression;][]{de2014spatial,satija2015spatial,berglund2018spatial}, and histopathology \citep{liao2021uncovering}. To bridge this gap, recently new spatially resolved transcriptomics (SRT) technologies have emerged, enabling gene expression profiling while maintaining high spatial characterization. These innovative technologies fall into two mainstreams: 1) imaging-based SRT platforms, including sequential single-molecule fluorescence \textit{in situ} hybridization (seqFISH) \citep{lubeck2014single}, multiplexed error-robust FISH (MERFISH) \citep{chen2015spatially}, and spatially-resolved transcript amplicon readout mapping (STARmap) \citep{wang2018three}; and 2) NGS-based SRT platforms, such as spatial transcriptomics (ST) \citep{staahl2016visualization}, 10x Visium ST, high-definition ST (HDST) \citep{vickovic2019high}, and Slide-seq \citep{rodriques2019slide}. The former are typically limited to dozens or hundreds of preselected genes, whereas the latter can reconstruct a transcriptome-wide spatial map covering expression levels of tens of thousands of protein-coding genes, providing a more comprehensive understanding. With these advancements, NGS-based SRT techniques have become pivotal in discovering novel insights in biomedical research. 

From a spatial statistics perspective, NGS-based SRT sequence count data can be viewed as a spatial map of marked points, where a multivariate count vector features each point. The high dimensionality of such data poses a challenge to standard statistical methods, leading to a long-standing interest in developing advanced clustering algorithms \citep{tadesse2005bayesian}. In the SRT context, this fundamental clustering task is aligned with identifying spatial domains, defined as distinct regions with similar molecular or morphological characteristics of clinical or biological significance. This step is the foundation for several downstream analyses, including but not limited to the spatial domain-based differential expression analysis, trajectory analysis, and functional pathway analysis \citep{thrane2018spatially,moses2022museum}. Directly applying those scRNA-seq clustering methods \citep{kiselev2019challenges}, such as the Louvain method \citep{blondel2008fast}, to SRT data often yields disjointed spatial domain patterns due to their reliance solely on gene expression levels. In contrast, recent advancements in SRT clustering approaches can be classified into two categories based on the additional information that they leverage to enhance clustering performance. The first category incorporates the geospatial profile by employing Markov random field (MRF) models to improve spatial coherence. Representative methods in this category include BayesSpace \citep{zhao2021spatial}, SC-MEB \citep{yang2022sc}, DR-SC \citep{liu2022joint}, and BayesCafe \citep{li2023interpretable}. The second category focuses on utilizing the paired histology image with SRT data. For instance, stLearn \citep{pham2020stlearn} extracts tissue morphological features, and SpaGCN \citep{hu2021spagcn} uses RGB color values from the raw histology image. In contrast, iIMPACT \citep{jiang2023integrating} leverages detailed morphological information from the artificial intelligence (AI)-reconstructed histology images. 

While the aforementioned methods have demonstrated their ability to produce continuous spatial domain patterns, they all possess clear limitations. Firstly, those methods typically involve transforming count data, which often contains numerous zeros, into normalized data for analysis. This preprocessing may not reflect the underlying data generation mechanism, potentially leading to a loss of statistical power \citep{sun2020statistical}. Furthermore, most methods require dimension reduction techniques, such as principal component analysis (PCA), t-distributed stochastic neighbor embedding (tSNE) \citep{van2008visualizing}, and uniform manifold approximation and projection (UMAP) \citep{becht2019dimensionality}. While effective against the curse of dimensionality, these techniques lack direct biological interpretability. Most importantly, all methods require the number of clusters $K$ to be known, or they utilize some criterion to determine the optimal $K$ in the post-processing step. However, an incorrectly specified $K$ can lead to misleading spatial domain patterns, compromising the effectiveness of downstream analyses.

In response to the limitations previously outlined and the motivation detailed in Section \ref{sec:data_motivate}, which highlights a strong correlation between spatial domain and gene expression patterns, we have developed a Bayesian nonparametric (BNP) spatial clustering framework, namely BNPSpace. This framework integrates four components into a hierarchical model. Firstly, it employs a multivariate count generating process based on the zero-inflated Poisson (ZIP) model for directly modeling SRT count data, thereby avoiding the need for \textit{ad hoc} data normalization methods. Secondly, following \cite{li2023interpretable}, BNPSpace adopts a feature selection strategy to provide low-dimensional representations of SRT data in terms of a list of discriminating genes (DGs), thus facilitating direct biological interpretability. Distinctively, BNPSpace clusters samples based on the similarity of their DG count vectors using a mixture of finite mixtures (MFM) model, allowing for an automatic selection of the number of spatial domains. Moreover, BNPSpace uses an MRF prior to accounting for the geospatial profile of SRT data, which in turn enhances clustering accuracy. To the best of our knowledge, BNPSpace uniquely fills a gap in BNP methodologies by jointly clustering samples with spatial context and selecting discriminating features, leveraging their mutual relationship. This comprehensive framework also offers a robust alternative in SRT clustering analysis that predominantly relies on normality assumptions and dimension reduction techniques. The efficacy of BNPSpace is demonstrated \textit{via} an extensive simulation study, exploring various spatial patterns and zero-inflation scenarios. The application of BNPSpace to the mouse olfactory bulb (MOB) ST data and human dorsolateral prefrontal cortex (DLPFC) 10x Visium data underscores its superiority. Notably, spatial domains identified by BNPSpace exhibit a more substantial alignment with manual annotations provided by experienced pathologists. Additionally, the associated DGs demonstrate biological relevance, substantiated by their significant overlap with DLPFC gene sets validated through real biological experiments.

The structure of this paper is organized as follows. We begin in Section~\ref{sec:data} by introducing the data structure and notations, complemented by an illustrative example in Section~\ref{sec:data_motivate}. Section~\ref{sec:meth} is dedicated to the formulation of the proposed BNP spatial clustering framework. In Section~\ref{sec:BI}, we overview the Markov chain Monte Carlo (MCMC) algorithms and briefly discuss the resulting posterior inference. In Sections~\ref{sim} and~\ref{app}, we evaluate the proposed model using simulated and real data, including benchmark MOB ST data and the human DLPFC 10X Visium data. Section~\ref{sec:conc} concludes the paper with a summary and a discussion of potential avenues for future research.

\section{Data Structure and Notations}\label{sec:data}

NGS-based spatial molecular profiling platforms, such as ST and the enhanced 10x Visium platforms, measure genome-wide expression levels encompassing over ten thousand genes across thousands of spatial locations referred to as `spots' on a tissue section (see an example in Figure \ref{fig:1}(a)). We denote the total number of spots by $n$ and the total number of genes by $p$, where typically $n\ll p$. The resulting SRT data can be summarized into two profiles: the molecular and geospatial profiles. The molecular profile is represented by an $n \times p$ count matrix $\bm{Y}$, with each entry $y_{ij} \in \mathbb{N}$ being the read count for gene $j$ observed in spot $i$. Here, we use $i=1,\ldots,n$ and $j=1,\ldots,p$ to index spots and genes, respectively. 

The corresponding geospatial profile is depicted by an $n \times 2$ matrix $\bm{T}$, where the $i$-th row $\bm{t}_{i\cdot}=(t_{i1},t_{i2})\in\mathbb{R}^2$ gives the x and y-coordinates of the $i$-th spot's center. We use the notation $\bm{x}_{i\cdot}$ and $\bm{x}_{\cdot j}$ to denote the $i$-th row and $j$-th column of any matrix $\bm{X}$, respectively, throughout the paper. Figure \ref{fig:1}(b) displays a representative example of the SRT molecular and spatial profiles analyzed in this paper. Notably, the $n$ spots are approximately arrayed on a two-dimensional square or triangular lattices, with each interior spot surrounded by four or six adjacent spots in the ST and 10x Visium platforms. This spatial configuration allows us to alternatively define the geospatial profile by an $n\times n$ binary adjacency matrix $\bm{G}$, where each entry $g_{ii'}=1$ if spot $i$ and $i'$ are neighbors and $g_{ii'}=0$ otherwise. We can derive $\bm{G}$ from $\bm{T}$ by evaluating the Euclidean distances between all spot pairs, i.e., $g_{ii'}={I}(\sqrt{(t_{i1}-t_{i'1})^2+(t_{i2}-t_{i'2})^2}<c_0)$, where ${I}(\cdot)$ denotes the indicator function and $c_0\in\mathbb{R}^+$ is the desired threshold. Note that all diagonal entries in $\bm{G}$ are set to zero by default, i.e., $g_{ii}=0,\forall i$. 

By introducing the above notations, an NGS-based SRT dataset can be conceptualized as a spatial map of marked points. Each point indexed by $i$ refers to a spot at location $\bm{t}_{i\cdot}$, characterized by multivariate discrete marks $\bm{y}_{i\cdot}$. The analysis of SRT data thus becomes an investigation of this marked point pattern.

\section{A Motivating Example}\label{sec:data_motivate}
This study is inspired by the analysis of spatially variable genes (SVGs) when analyzing the MOB ST data collected by the original ST platform  \citep{staahl2016visualization}. The data are openly accessible through the Spatial Research Lab \footnote{https://www.spatialresearch.org/resources-published-datasets/doi-10-1126science-aaf2403/}. Among the 12 available replicates, we choose replicate $11$ for illustration, which encompassed $16,034$ genes across $282$ spots arranged in a square lattice. We followed the quality control procedure recommended by \cite{li2023interpretable}, eliminating spots with total read counts below $100$ and genes where the proportion of zeros exceed $90\%$ and the maximum read counts were below $10$. This process yielded a refined dataset comprising $n=278$ spots and $p=1,117$ genes. The detection of SVGs is a fundamental question in the analysis of SRT data, testing the association between a gene's expression levels $\bm{y}_{\cdot j}$ and spatial coordinates $\bm{T}$ against a null hypothesis of spatial invariance. Biologically, SVGs reflect the tissue heterogeneity and underlying tissue structure that drive the differentiated expression across different spatial domains. Thus, SVGs are potentially significant and may lead to novel biological insights.

\begin{figure}
\begin{center}
\includegraphics[width = 1\textwidth]{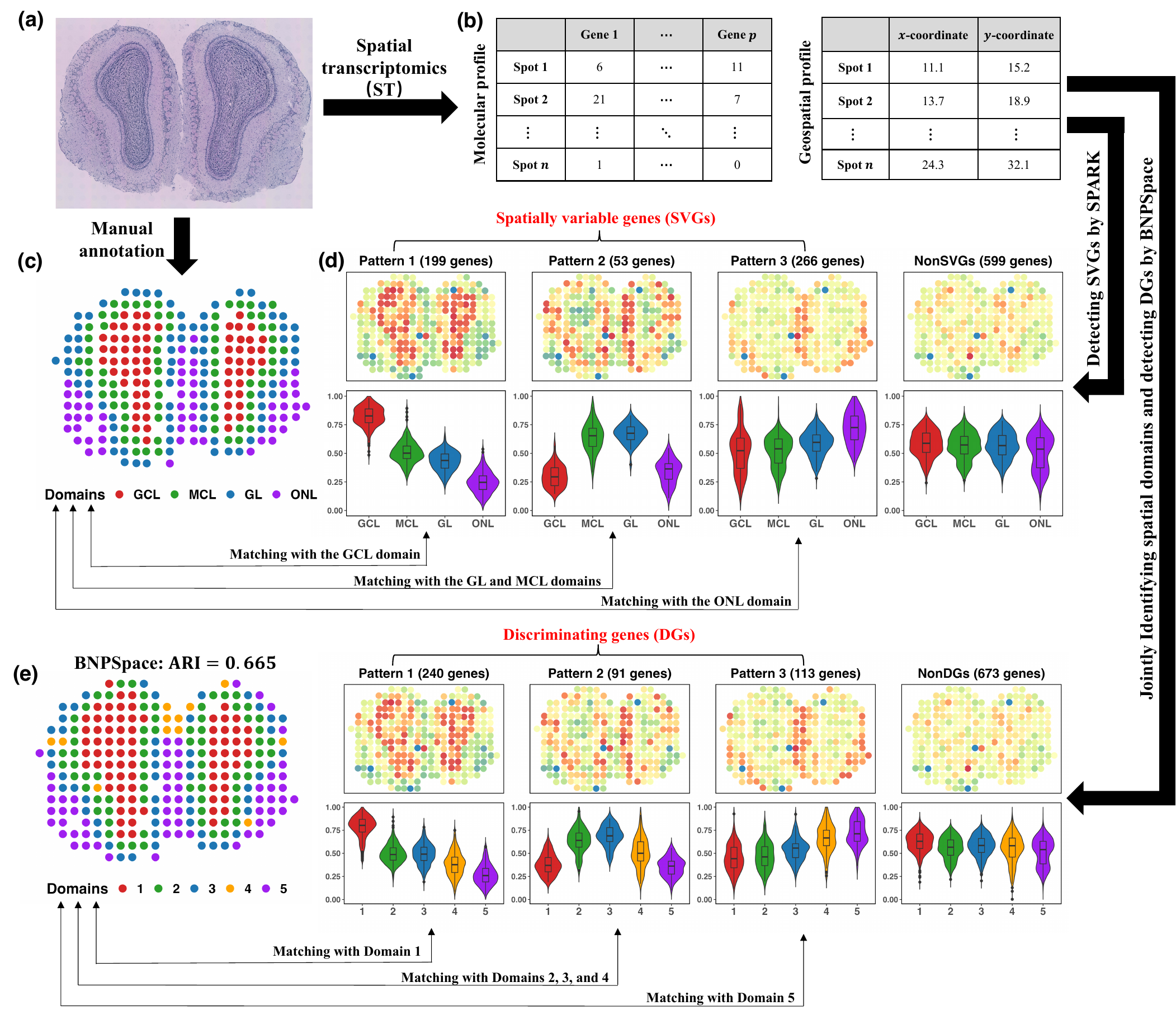}
\end{center}
\caption{MOB ST data analysis: (a) Paired histology image; (b) Molecular and geospatial profiles; (c) Manually annotated spatial domains; (d) Three distinct spatial expression patterns derived from $518$ SPARK-identified SVGs by hierarchical clustering, along with the average expression patterns of the $599$ nonSVGs; (e) BNPSpace-identified spatial domains, three distinct spatial expression patterns derived from $444$ BNPSpace-identified DGs, along with the averaged expression patterns of the remaining $673$ nonDGs.
 \label{fig:1}}
\end{figure}

In our analysis of the MOB ST data, we first employed SPARK  \citep{sun2020statistical}, a Gaussian process-based geostatistical model, to detect SVGs among all $1,117$ candidate genes. This approach yielded $520$ SVGs, while controlling a false negative rate (FDR) of $5\%$ \citep{benjamini1995controlling}. Based on the four manually annotated layers (see Figure \ref{fig:1}(c)), which served as the ground truth, we performed a Kruskal-Wallis test \citep{kruskal1952use} for each SVG with a null hypothesis that the medians of relative gene expression among the four layers are equivalent. It turned out that $457$ genes, accounting for nearly $90\%$ of the SPARK-identified SVGs, were statistically significant at an FDR threshold of $5\%$. This finding underscores a strong linkage between the spatial domain profile and SVG expression patterns. To confirm this association, we followed \cite{sun2020statistical} to conduct agglomerative hierarchical clustering on all $520$ SVGs based on their relative expression levels, leading to three distinctive mean expression patterns as shown in Figure \ref{fig:1}(d). Their corresponding violin plots, which display the distributions of relative expression levels across the four layers, clearly suggest that those $199$, $53$, and $266$ SVGs belonging to Patterns 1, 2, and 3 were highly expressed in the granule cell layer (GCL), mitral cell and glomerular layers (GL/MCL), and nerve layer (ONL), respectively. In contrast, the mean relative expression of those $599$ nonSVGs, visualized in Figure \ref{fig:1}(d), fails to exhibit any distinguishable patterns aligned with the ground truth.

These findings imply that certain genes, notably SVGs, play a pivotal role in distinguishing the $n$ spots into mutually exclusive spatial domains. In turn, a well-defined spatial domain profile can be instrumental in pinpointing genes with biologically meaningful spatial patterns. Although current methods for SVG detection and spatial domain identification have individually contributed to the analysis of SRT data, the mutual relationship between SVGs and spatial domains has not been fully leveraged in the literature. Additionally, both methods typically require \textit{ad hoc} data normalization steps at the spot or gene level, which may not fully accommodate the characteristics of SRT data and thus compromise statistical power. Hence, a joint modeling framework tailored to address the high-dimensionality, zero-inflation, and over-dispersion inherent in NGS-based SRT data could simultaneously refine the identification of spatial domains and biologically relevant genes.

\section{Model}
\label{sec:meth}
Here, we introduce the BNP spatial clustering framework, which incorporates a feature selection mechanism. The graphical and hierarchical formulations of the proposed model are depicted in Figures S1 and Table S1, respectively, in the supplement.

\subsection{Modeling Raw Counts \textit{via} a ZIP Model}\label{sec:ZIP}
NGS count data, including scRNA-Seq and SRT data, usually suffers from zero-inflation, particularly when sequencing depth is inadequate \citep{risso2018general,li2021bayesian}. For instance, the aforementioned MOB ST data exhibits a high sparsity, with $72\%$ of read counts being zeros. Zero-inflation typically arises from biological reasons, such as cell subpopulations and transient states where a gene was not expressed, as well as technical factors like dropouts, where a gene, though expressed, remains undetected due to limitations in the sequencing process. Prior research has demonstrated that computational methods that account for the high proportion of zeros tend to improve both model fitting and the accuracy of differential gene expression detection \citep{finak2015mast,l2016pooling}. Therefore, our modeling framework begins with the implementation of a ZIP model to characterize read counts with excess zeros:
\begin{equation} \label{model_1}
y_{ij}|\pi_i,\mu_{ij} \sim \pi_{i} {I}(y_{ij}=0) + (1- \pi_{i})\int \text{Poi}(s_{i}\mu_{ij}) {H}(d\mu_{ij}),\text{ for } i=1,\ldots,n,j=1,\ldots,p,
\end{equation} 
where $\pi_i\in[0,1]$ represents the underlying proportion of extra zeros (also known as false or structural zeros) at spot $i$. The indicator function, denoted as ${I}(\cdot)$, constrains the first kernel in (\ref{model_1}) to be degenerate at zeros, thereby allowing for zero-inflation. The general mixing distribution, denoted as  ${H}(\cdot)$, model the over-dispersion of the SRT data. When ${H}(\cdot)$ is a gamma distribution, the above model defines a zero-inflated negative binomial (ZINB) model \citep{li2017bayesian}. Alternatively, we can rewrite this ZIP model by introducing an $n\times p$ latent binary matrix $\bm{R}$, with each entry $r_{ij}|\pi_i\sim\text{Bern}(\pi_i)$, such that if $r_{ij}=1$, then $y_{ij}=0$, whereas if $r_{ij}=0$, then $y_{ij}|\mu_{ij}\sim\int \text{Poi}(s_{i}\mu_{ij}) {H}(d\mu_{ij})$. This independent Bernoulli prior assumption can be further relaxed by placing a beta hyperprior on $\pi_i$, i.e., $\pi_i\sim\text{Be}(\alpha_\pi,\beta_\pi)$, leading to a beta-Bernoulli prior on $r_{ij}$ with an expectation of $\alpha_\pi/(\alpha_\pi+\beta_\pi)$. Setting $\alpha_\pi= \beta_\pi=1$ results in a uniform prior on $\pi_i$.

We decompose the mean of the Poisson distribution into a product of the size factor $s_{i} \in \mathbb{R}^{+}$, which adjusts for spot-specific effects, and the normalized gene expression level $\mu_{ij} \in \mathbb{R}^{+}$. Such a multiplicative formulation of the Poisson mean is typical in both the frequentist \citep{10.1214/11-AOAS493, li2012normalization, cameron2013regression} and the Bayesian \citep{banerjee2003hierarchical, airoldi2016improving} literature to accommodate latent heterogeneity and over-dispersion in count data. Although the size factor parameters, $\bm{s}=(s_1,\ldots,s_n)^\top$, can be estimated through a Dirichlet process mixture (DPM) prior model under mean constraints \citep{li2017bayesian,li2019bayesian}, a simple and practical approach is to set each size factor $s_{i}$ proportional to the total sum of counts in the corresponding spot, i.e., $s_{i} \propto \sum_{j}y_{ij}$ \citep{sun2020statistical}, with a constraint $\prod_{i=1}^{n} s_{i} = 1$ to ensure identifiability \citep{li2021bayesian}. This leads to $s_{i} = \sum_{j=1}^{p}y_{ij} / \sqrt[n]{\prod_{i=1}^{n} \sum_{j=1}^{p}y_{ij}}$. By accounting for zero-inflation through the extra zero indicators $\bm{R}$, over-dispersion through the mixing distribution ${H}(\cdot)$, and sample heterogeneity through the size factors $\bm{s}$, the ZIP model yields a denoised version of gene expression counts $\bm{Y}$, denoted as an $n\times p$ matrix $\bm{M}$, where each entry $\mu_{ij}$ is the latent normalized expression levels for gene $j$ at spot $i$.

\subsection{Detecting DGs \textit{via} a Feature Selection Scheme}\label{sec:feature}
In the context of clustering high-dimensional data, such as the SRT data generated by NGS-based SRT platforms, it becomes evident that numerous genes (e.g., nonSVGs) contribute minimal information for clustering purposes, as illustrated in our motivating example from Section \ref{sec:data_motivate}. The inclusion of such genes can not only complicate the clustering process but also impede the identification of true clusters. Thus, we envision that only a subset of measured genes is relevant to discriminate $n$ spots into distinct clusters, namely DGs. Unlike other SRT clustering approaches that rely on low-dimensional data representations, we propose a dimensional reduction approach \textit{via} the selection of DGs. This strategy provides a direct biological interpretation and preserves substantial information inherent in the original data, thereby enhancing the overall model robustness.

To identify DGs, we introduce a latent binary vector  $\bm{\gamma} = (\gamma_{1}, \ldots, \gamma_{p})^{\top}$, where $\gamma_{j} = 1$ indicates that gene $j$ is differentially expressed across spatial domains, and $\gamma_{j} = 0$ otherwise. Our model formulation assumes that read counts $y_{ij}$'s from a DG (i.e., $\gamma_j=1$) follows a ZIP mixture distribution, whereas those from a nonDG (i.e., $\gamma_j=0$) are modeled with a ZIP characterized by a common mean parameter $\mu_{0j}$, representing a `null' model of no gene expression heterogeneity. Conditioning on the DG indicator, we can rewrite (\ref{model_1}) as follows,
\begin{equation} \label{model_2}
y_{ij} | r_{ij}, \mu_{ij},\mu_{0j},\gamma_j\sim
\begin{cases}
r_{ij}I(y_{ij}=0)+(1-r_{ij})\int\text{Poi}\left(s_{i}\mu_{ij}\right){H}(d\mu_{ij}) & \text{ if } \gamma_j=1\\
r_{ij}I(y_{ij}=0)+(1-r_{ij})\text{Poi}\left(s_{i}\mu_{0j}\right) & \text{ if } \gamma_j=0
\end{cases}.
\end{equation}

A product of Bernoulli distribution with the same parameter is a natural choice for $\bm{\gamma}$,
\begin{equation}\label{prior_gamma}
	\bm{\gamma}|\omega \sim \prod_{j=1}^p\text{Bern}(\omega).
\end{equation}
 Let $p_{\gamma} = \sum_{j=1}^{p}\gamma_{j}$ be the total number of DGs, with the prior equivalent to $p_{\gamma} | \omega \sim \text{Bin}(p, \omega)$. The hyperparameter $\omega$ can be elicited as the proportion of DGs expected \textit{a priori}. To further enhance the model's robustness against variations in $\omega$, we assign a beta distribution to it, $\omega \sim \text{Be}(\alpha_{\omega},\beta_{\omega})$, leading to a beta-binomial prior distribution on $p_{\gamma}$. We follow \cite{tadesse2005bayesian} to employ a vague prior for $\omega$ by setting $a_\omega+b_\omega=2$, thereby allowing the data to drive the estimation of the proportion of DGs.

\subsection{Clustering Spots \textit{via} the MFM Model}\label{sec:MFM}
In this section, we explore the application of the MFM model in the context of SRT clustering analysis. Our model assumes that clustering of each spot is primarily driven by the read counts from DGs, as represented by $\bm{y}_{i\cdot}^{(\gamma)}$. Here, the superscripts $(\gamma)$ and $(\gamma^c)$ index the sets of DGs and nonDGs characterized by $\gamma_j=1$ and $0$, respectively. We introduce a latent clustering allocation vector, $\bm{z}=(z_1,\ldots,z_n)^\top$, where $z_i=k$ indicates the observation vector $\bm{y}_{i\cdot}^{(\gamma)}$ belongs to cluster $k$, thereby locating spot $i$ within spatial domain $k$, for some positive integer $k\in\mathbb{N}\backslash\{0\}$. Under a conjugate setup with $H(\cdot)$ as a gamma distribution,  we reformulate (\ref{model_1}) as follows:
\begin{equation} \label{model_3}
	y_{ij} | r_{ij}, \mu_{kj}^*,\mu_{0j},\gamma_j,z_i=k\sim
	\begin{cases}
		r_{ij}I(y_{ij}=0)+(1-r_{ij})\text{Poi}\left(s_{i}\mu_{kj}^*\right)& \text{ if } \gamma_j=1\\
		r_{ij}I(y_{ij}=0)+(1-r_{ij})\text{Poi}\left(s_{i}\mu_{0j}\right) & \text{ if } \gamma_j=0
	\end{cases},
\end{equation}
where $\mu_{kj}^*\sim\text{Ga}(a_\mu,b_\mu)$ represents the normalized expression level of gene $j$ as a DG, common to all none extra-zero read counts in cluster $k$. By introducing $\bm{\gamma}$ and $\bm{z}$ into our model, we establish an effective constraint on the latent normalized gene expression levels $\bm{M}$, which reflects the underlying biological processes. Specifically, each entry $\mu_{ij}=\mu_{kj}^*$ if $z_i=k$ and $\gamma_j=1$ while $\mu_{ij}=\mu_{0j}$ if $\gamma_j=0$.

Choosing priors for the clustering allocation vector $\bm{z}$ leads to different modeling approaches. For instance, a ZIP-based finite mixture model (FMM), as described in  \cite{li2023interpretable}, is achieved with a fixed number of clusters $K$ and a Dirichlet-multinomial prior configuration, i.e., $z_i|\bm{q}\sim\text{Multi}(1,\bm{q})$ and $\bm{q}\sim\text{Dir}(\bm{\alpha}_0)$. In this setup, $\bm{q}=(q_1,\ldots,q_K)^\top$ represents the cluster mixing proportions and $\bm{\alpha}_0=(\alpha_0,\ldots,\alpha_0)^\top$ is the concentration parameter for a symmetric Dirichlet distribution. Alternatively, considering $z_i|\cdot\sim\sum_{k=1}^\infty q_kI(z_i=k)$ with $q_k$ defined by the stick-breaking construction \citep{sethuraman1994constructive} results in a ZIP-based Dirichlet process mixture model (DPMM) \citep{antoniak1974mixtures,li2017bayesian}. This model can also be viewed by extending a $K$ component-FMM as $K\rightarrow\infty$ and employing a Dirichlet prior $\bm{q}\sim\text{Dir}(\bm{\alpha}_0/K)$. By integrating out the mixing proportions $\bm{q}$, we can obtain the P\'{o}lya Urn scheme for the clustering assignments $\bm{z}$, known as the Chinese restaurant process (CRP). The corresponding conditional distribution for each $z_i$ \citep{blackwell1973ferguson} is given as: 
\begin{equation} \label{polya_1}
	P\left(z_i=k | \bm{z}_{-i}\right) \propto \begin{cases}
		n_{k,-i} & \text { at an existing cluster } k \\
		\alpha_0 & \text { if } c \text{ is a new cluster }
	\end{cases},
\end{equation}
where $\bm{z}_{-i}$ denotes all entries in $\bm{z}$ excluding the $i$-th one and $n_{k,-i}$ is the number of spots in cluster $k$ ignoring the $i$-th one. This modeling framework allows us to cluster spots based on the expression levels of those DGs, by allowing $\mu_{ij}=\mu_{kj}^*$ when $r_{ij}=0$, for some $k$, while the number of clusters $K$ is estimated as a by-product of the usual posterior inference. 

While the CRP has a very attractive feature of simultaneous estimation of the
number of clusters and the cluster configuration and corresponding densities, a significant shortcoming of
this model was recently discovered \citep{miller2018mixture}. Even when the sample size approaches
infinity, the estimation of the {\em number of clusters} is inconsistent due to  extraneous clusters in the posterior of CRP. To alleviate this problem, \citet{miller2018mixture} proposed the MFM model, treating $K$ as a random variable and employing a truncated Poisson prior on it. Specifically, this prior setting can be summarized as
$z_{i} | K, \bm{q} \sim \sum_{k=1}^{K} q_KI(z_i=k)$, $\bm{q} | K \sim \text{Dir}(\bm{\alpha}_0)$, and $K-1 \sim \text{Poi}(\lambda)$, leading to a modified P\'{o}lya Urn scheme on the prior distribution of $\bm{z}$ as follows:
\begin{equation}\label{polya_2}
    P\left(z_i=k | \bm{z}_{-i},\right) \propto\begin{cases}
{n_{k,-i}+\alpha_0} & \text { at an existing cluster } k \\
\frac{V_n\left(|\bm{z}_{-i}|+1\right)}{ V_n\left(|\bm{z}_{-i}|\right)} \alpha_0 & \text { if } k \text { is a new cluster}
\end{cases},
\end{equation}
where $V_{n}(\cdot)$ is a pre-computed coefficient detailed in Section S1 of the supplement and $|\bm{z}_{-i}|$ denotes the number of unique values in $\bm{z}_{-i}$. Compared to the original P\'{o}lya Urn scheme, the modified version limits the likelihood of forming extraneous clusters by implementing a scaling factor ${V_{n}(\cdot+1)}/{V_{n}(\cdot)}$ that is less than one for new clusters.

To complete the model, we set the hyperparameters controlling the base measure $H(\cdot)$ to $\alpha_\mu=\beta_\mu=1$ for a mean and variance of $1$ in the gamma distribution. Additionally, we specify the concentration parameter $\alpha_0=1$ for a weakly informative setting and set $\lambda=1$ to favor a smaller number of clusters $K$ \textit{a priori}. This choice has been validated by \cite{miller2018mixture,geng2019probabilistic,geng2021bayesian} to guarantee consistency for the mixing distribution and the number of clusters.

\subsection{Integrating the Geospatial Profile  \textit{via} the MRF Model}\label{sec:MRF-MFM}
To take advantage of the additional spatial structure of NGS-based SRT data (i.e., the gene expression levels are measured on a lattice grid), numerous statistical models adopt an MRF prior on $\bm{z}$ to enhance the spatial coherence of neighboring spots \citep[see, e.g.,][]{zhu2018identification,liu2022joint,yang2022sc,jiang2023integrating,li2023interpretable}. This modeling approach is particularly effective with the FMMs, where the number of clusters $K$ is predetermined. Under this framework, the conditional distribution for each $z_i$ can be expressed as:
\begin{equation}\label{prior_mrf}
    P(z_{i} = k | \bm{z}_{-i}) \propto \exp\left(e_k+d \sum_{i'=1}^ng_{ii'}{I}(z_{i'} = k)\right), 
\end{equation}
where $e_k\in\mathbb{R}$ signifies the underlying mixing proportion of cluster $k$ and  $d\in\{0\}\cup\mathbb{R}^+$ controls the spatial dependency strength or spatial homogeneity among neighboring spots. Larger values of $d$ lead to high homogeneity in the clustering analysis, resulting in a  smooth spatial domain pattern. Note that if a spot does not have any neighbors, the configuration reduces to $z_i|\bm{q}\sim\text{Multi}(1,\bm{q})$, with each $q_k=\exp(e_k)/\sum_{k=1}^K\exp(e_k)$. 

A key limitation of applying the above MFM model directly to SRT data analysis is its neglect of the geospatial profile. Consequently, the resulting clusters do not incorporate spatial constraints. Integrating MRF with the MFM model which enables spatial clustering under a BNP framework is proposed to address this weakness \citep{hu2023bayesian,zhao2023spatial}. The priors of the MRF-constrained MFM model can be described as follows:
\begin{equation}\label{prior_z}
     P(\bm{z} |  \bm{q}) \propto\left(\prod_{i=1}^n q_{z_i}\right) \exp \left(d \sum_{i<i'} g_{ii'} {I}(z_i=z_{i'}) \right), \quad   \bm{q} | K \sim \operatorname{Dir}(\bm{\alpha}_0), \quad K-1 \sim \operatorname{Poi}(\lambda). 
\end{equation}
As established in Theorem \ref{thm1}, a notable aspect of this constrained model is that the MRF constraints affect only the finite component of the MFM model. The full proof of this theorem is provided in Section S1 of the supplement.
\begin{theorem}\label{thm1}
Under the MRF-constrained MFM model specification in Equation (\ref{prior_z}), the P\'{o}lya Urn Scheme is represented by:
\begin{equation} \label{polya_3}
    P\left(z_i=k \mid \bm{z}_{-i}\right) \propto\begin{cases}
{(n_{k,-i}+\alpha_0)\exp \left(d \sum_{i'=1}^n g_{ii'} I\left(z_{i'}=k\right)\right)} & \text{at an existing cluster  } k \\
\frac{V_n\left(|\bm{z}_{-i}|+1\right) }{ V_n\left(|\bm{z}_{-i}|\right)} \alpha_0 & \text {if } k \text { is a new cluster }
\end{cases},
\end{equation}
where all notations above are defined earlier in the paper.
\end{theorem}

P\'{o}lya urn scheme in Theorem \ref{thm1} will let nearby spots have a higher
probability being clustered together when $d > 0$.
This will enforce the locally contiguous clusters.
The globally discontiguous clusters will be learned from the data itself. The proposed model \eqref{prior_z} will reduce to traditional MFM model when $ d=0$. Compared with traditional spatial cluster algorithms, a distinction of the proposed method is its ability to
capture both locally contiguous clusters and globally discontiguous
clusters.
The selection of the hyperparameter $d$ in \eqref{polya_3} is important due to its role in controlling spatial smoothness. While a higher $d$ value encourages a smoother spatial domain pattern, it may lead to a phase transition problem. We treat $d$ as a fixed hyperparameter, following \cite{li2010bayesian} and \cite{stingo2013integrative}. To choose an suitable value of $d$, we recommend using the penalized Bayesian information criteria (pBIC) \citep{pan2007penalized}, detailed in Section S3 of the supplement.

\section{Model Fitting}
\label{sec:BI}

Our study focuses on detecting DGs through the DG indicator vector $\bm{\gamma}$ and identifying spatial domains through the cluster allocation vector $\bm{z}$. We aim to sample from the posterior distributions of the unknown parameters $\bm{M}$, $\bm{R}$, $\bm{\pi}$, $\bm{\gamma}$, and $\bm{z}$. The data likelihood is
\begin{equation}
    L(\bm{Y} | \bm{R}, \bm{M}, \bm{\gamma}, \bm{z}) = \prod_{i=1}^{n} \prod_{j = 1}^{p} P(y_{ij} \mid r_{ij}, \gamma_{j}, z_{i}, \mu_{ij}),
\end{equation}
where $P(y_{ij} \mid r_{ij}, \gamma_{j}, z_{i}, \mu_{ij})$ is the likelihood of Equation (\ref{model_3}), which we detailed in Section S2. 
We employ an MCMC algorithm that integrates Metropolis search variable selection \citep{george1997approaches, brown1998multivariate} and a collapsed Gibbs sampler \citep{neal2000markov} to generate posterior samples for each parameter in the MRF-constrained MFM model. The proposed algorithm uses an efficient collapsed Gibbs sampler that analytically
marginalizes out~$K$. Detailed descriptions of the MCMC procedure are provided in Section S2 of the supplement. Posterior inference on the relevant parameters is achieved \textit{via} post-processing MCMC samples after the burn-in period, the details of which are elaborated below.

\subsection{Posterior Inference on DGs}
For the selection of DGs, we can rely on the \textit{maximum a posteriori} (MAP) estimate of $\bm{\gamma}$,
\begin{equation}
	\hat{\bm{\gamma}}^{\text{MAP}}   =  \underset{\bm \gamma}{\mathrm{argmax}}~ L\left(\bm{Y}|\bm{R}^{(u)},\bm{M}^{(u)},\bm{\gamma}^{(u)},\bm{z}^{(u)}\right)\pi\left(\bm{\gamma}^{(u)}\right),
\end{equation}
where the superscript $(u)$ index the MCMC iterations post-burn-in, and $U$ is the total iteration number. To comprehensively summarize $\bm{\gamma}$, we focus on their marginal posterior probabilities of inclusion (PPI), where $\text{PPI}_j\approx\sum_{u=1}^U\gamma_j^{(u)}/U$. 
Subsequently, DGs are identified based on whether their PPI values exceed a specific threshold $c\in[0,1]$:
\begin{equation}
		\bm{\hat{\gamma}}^{\text{PPI}} =  \Big({I}(\text{PPI}_1\geq c),\cdots,{I}(\text{PPI}_p\geq c)\Big)^\top.
\end{equation}
To control for multiple testing, we adopt a thresholding approach \citep{newton2004detecting} that ensures the expected Bayesian FDR (BFDR) below a predetermined level (e.g., $5\%$). The BFDR is computed as follows:
\begin{equation}
	\text{BFDR}(c)  =  \frac{\sum_{j=1}^p(1-\text{PPI}_j){I}(1-\text{PPI}_j<c)}{\sum_{j=1}^p{I}(1-\text{PPI}_j<c)}
\end{equation}
where BFDR($c$) is the desired significance level.  According to \cite{peterson2015bayesian}, a cutoff of $c = 0.5$ provides a reasonable BFDR balance; therefore, we set $c = 0.5$, referred to as the {\it{median}} model. 

\subsection{Posterior Inference on Spatial Domains}
For summarizing the posterior distribution of $\bm{z}$, we can also use the MAP estimate,
\begin{equation}
	\hat{\bm{z}}^{\text{MAP}}   =  \underset{\bm z}{\mathrm{argmax}}~ L\left(\bm{Y}|\bm{R}^{(u)},\bm{M}^{(u)},\bm{\gamma}^{(u)},\bm{z}^{(u)}\right)\pi\left(\bm{z}^{(u)}|\bm{q}^{(u)}\right)
\end{equation}
Additionally, we may obtain a summary of $\bm{z}$ based on the pairwise probability matrix (PPM) \citep{dahl2006model}. The PPM, an $n\times n$ symmetric matrix, calculates the posterior pairwise probabilities of co-clustering; that is, the probability that spot $i$ and $i'$ are assigned to the same cluster: $\text{PPM}_{i,i'}\approx\sum_{u=1}^U{I}(z_i^{(u)}=z_{i'}^{(u)})/U$. A point estimate $\hat{\bm z}^{\text{PPM}}$ can then be derived by minimizing the sum of squared deviations between its association matrix and the PPM:
\begin{equation}
		\hat{\bm z}^{\text{PPM}}  =  \underset{\bm z}{\mathrm{argmax}}~\sum_{i<i'} \big({I}({z}_i={z}_{i'})-\text{PPM}_{ii'}\big)^2.
\end{equation}
The PPM estimate has the advantage of utilizing information from all clusterings through the PPM. It is also intuitively appealing because it selects the `average' clustering rather than forming a clustering \textit{via} an external, \textit{ad hoc} clustering algorithm. This approach ensures a more comprehensive and representative summary of the clustering outcomes.

\section{Simulation}
\label{sim} 

In this section, we provide an overview of the simulation studies, with a comprehensive description presented in Section S4 of the supplement.

\begin{figure}
\begin{center}
\includegraphics[width = 1\textwidth]{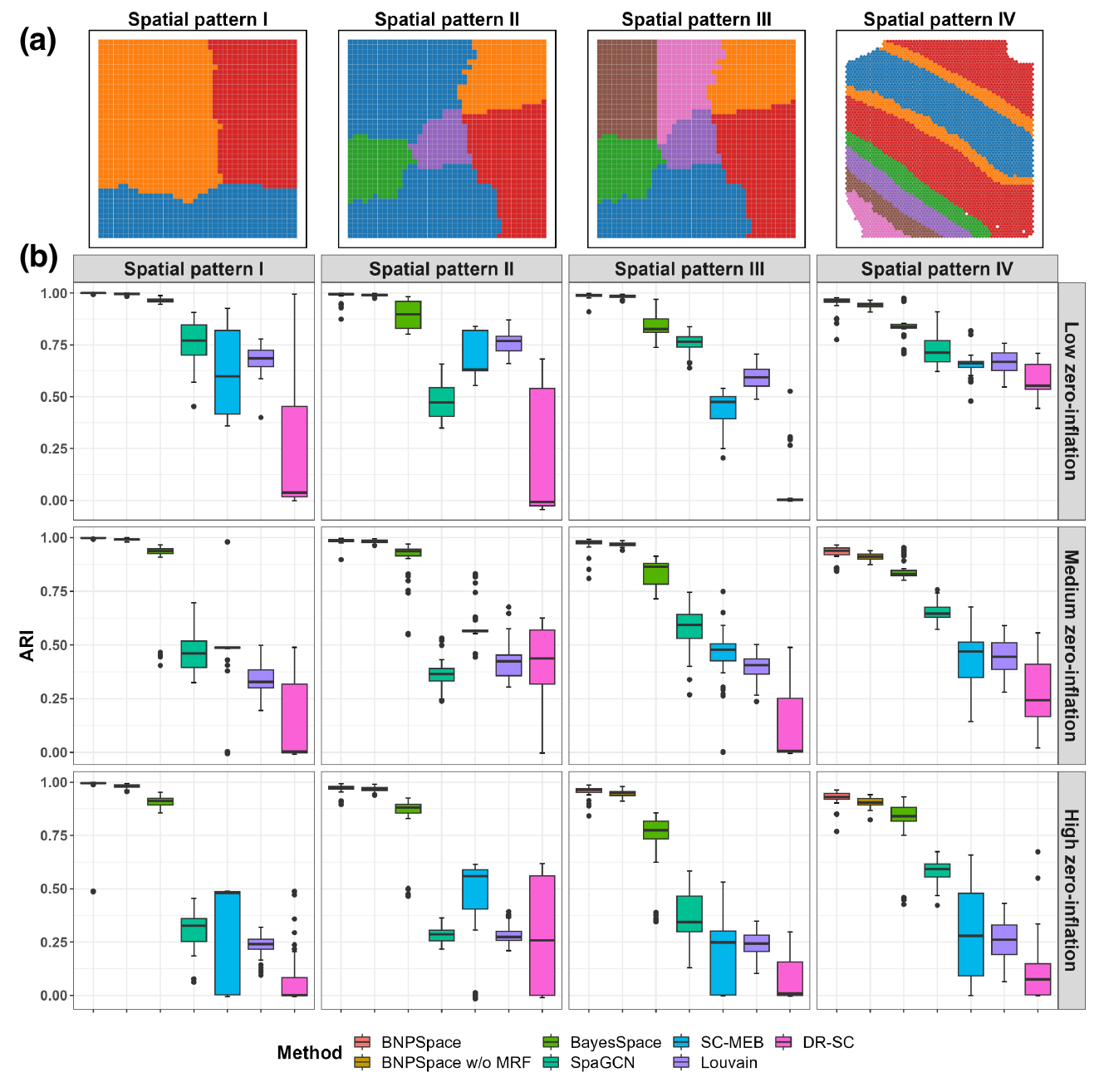}
\end{center}
\caption{Simulated data analysis: (a) Four spatial domain patterns; (b) Boxplots of ARIs achieved by BNPSpace and competing methods across various scenarios in terms of spatial domain patterns and zero-inflation settings. \label{fig:2}}
\end{figure}

We generated simulated SRT datasets based on three artificial spatial patterns and one real spatial domain pattern from the human DLPFC 10x Visium data. The first three patterns, displayed in Figure \ref{fig:2}(a) as Patterns I, II, and III, are on an $n=40\times40$ square lattice. Their $\bm{z}$'s were generated from Potts models with $K=3$, $5$, and $7$ states, respectively, using the \verb|sampler.mrf| function in the \verb|R| package \verb|GiRaF| and applying the same smoothing parameter. The real spatial pattern, illustrated as Pattern IV in Figure \ref{fig:2}(a), contains $n=4,788$ spots across $K=7$ spatial domains. Conditional on these four spatial patterns represented by $\bm{z}$, we followed the data generative scheme described in Section S4.1 to generate a count table $\bm{Y}$. For the choice of the extra zero proportions $\pi_i$, we randomly selected $10\%$, $20\%$, or $30\%$ of the entries in $\bm{Y}$ and set their values to zero. The above process, combined with the four spatial patterns and three zero-inflation settings, resulted in $4\times3=12$ different scenarios. For each scenario, we generated $50$ data replicates.

We utilized the aforementioned $12\times50=600$ simulated SRT datasets with $p=1000$ genes, $20$ of which are true DGs, to assess the performance of BNPSpace against a variety of alternative methods surveyed in Section \ref{sec:intro}. For approaches that require a low-dimensional representation of SRT data, including Louvain, BayesSpace, SC-MEB, and DR-SC, we selected a latent space dimension of $15$ to ensure maximum retention of information for clustering analysis. For those methods lacking a built-in mechanism for automatically selecting $K$, such as Louvain, BayesSpace, and SpaGCN, we fixed $K$ to its true value. In contrast, SC-MEB and DR-SC employed a modified Bayesian Information Criterion to determine $K$ \citep{ma2017concave}. To assess the role of spatial information, we also ran BNPSpace without the MRF prior (or equivalently, by setting $d=0$). We implemented BNPSpace following the default prior specifications and algorithm settings as elaborated in the paper. Further information can be found in Section S4.2 of the supplement. 

To evaluate the clustering performance through $\bm{z}$ across various methods, we used the adjusted Rand index \citep[ARI,][]{santos2009use}. The ARI, which ranges from 0 to 1, measures the similarity between two different partitions. Higher ARI values indicate more accurate clustering outcomes, with a value of one indicating a perfect match. In assessing the efficacy of identifying DGs \textit{via} the binary vector $\bm{\gamma}$, we calculated the area under the receiver operating characteristic curve (AUC). AUC incorporates both sensitivity and specificity across different PPI thresholds. Given that DGs are typically a tiny fraction of the total, we also computed the Matthews correlation coefficient (MCC), pinpointing a specific PPI threshold. For both AUC $\in[0,1]$ and MCC $\in[-1,1]$, the larger the index, the more accurate the inference. Rigorous definitions of all these performance metrics are provided in Section S4.3 of the supplement.

The main graphical summaries on $\bm{z}$ are presented in Figure \ref{fig:2}(b) and Figure S3, whereas the numerical results related to $\bm{\gamma}$ are summarized in Table S2 and S3 of the supplement. Based on these results, we draw the following conclusions: 1) BNPSpace exhibited superior clustering effectiveness, achieving an ARI value close to one across all patterns and zero-inflation settings; 2) An increase in data sparsity led to greater disparity between BNPSpace and other methods, underscoring the necessity of modeling extra zeros; 3) BNPSpace consistently identified the correct number of clusters with a success rate exceeding $85\%$ among all $50$ replicates under each scenario; and 4) BNPSpace accurately identified all true DGs with an average AUC $> 0.99$, as shown in Table S1 and S2. Furthermore, we compared the performance of BNPSpace with and without the MRF component. Through the Wilcoxon signed rank test, we obtained a maximum $p$-value of $4.62 \times 10^{-4}$ across all scenarios, suggesting that incorporating spatial information enhanced the clustering performance.

\section{Application}
\label{app}

\subsection{Application to the MOB ST Data}
\label{MOB}
We first applied BNPSpace to analyze the preprocessed MOB ST data introduced in Section \ref{sec:data_motivate}, using the same prior specifications and algorithm settings as in the simulation study. For the MRF hyperparameter, we set $d=1$ guided by the pBIC plot in Figure S4(a) of the supplement. To diagnose MCMC convergence, we ran three independent MCMC chains with diverse initializations. The trace plots for the ARI, the number of DGs, and two randomly selected parameters in $\bm{M}$ are presented in Figures S5 of the supplement and indicate satisfactory convergence. Additionally, to ensure the robustness of the feature selection outcomes, we calculated pairwise Pearson correlation coefficients of PPIs among these chains, with coefficients ranging from $0.893$ to $0.951$, further affirming good convergence. We then aggregated the outputs of all chains for posterior inference. For comparative methods, we employed their standard settings and set the number of spatial domains $K=4$. The computational efficacy is detailed in Table S4 of the supplement. 

\begin{figure}
\begin{center}
\includegraphics[width = 1\textwidth]{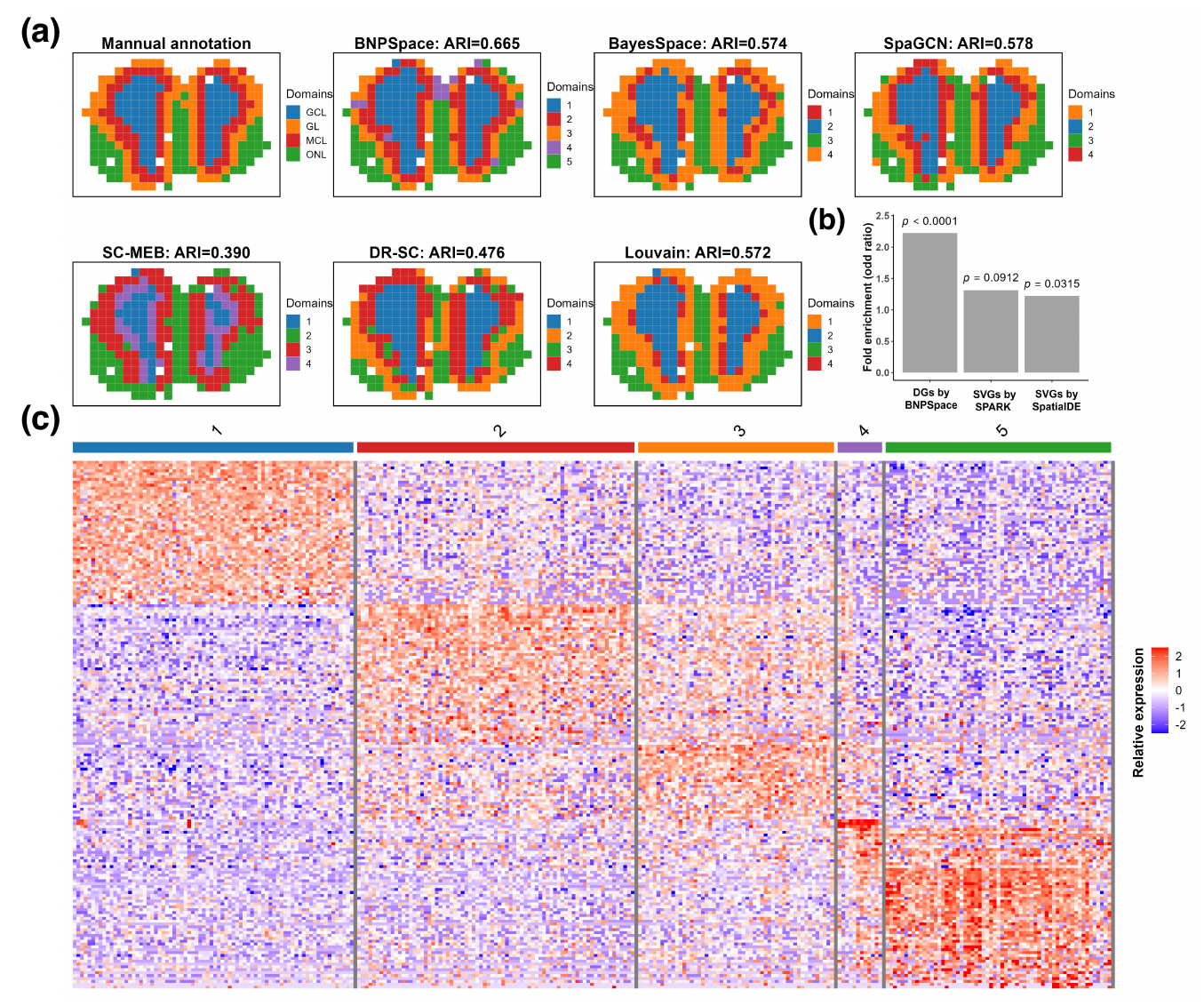}
\end{center}
\caption{MOB ST data analysis: (a) Spatial domains annotated by pathologists and identified by BNPSpace and competing methods; (b) Gene enrichment analysis between $444$ BNPSpace-identified DGs and $572$ SPARK-identified SVGs and $2109$ SpatialDE-identified SVGs; (c) Heatmap of $444$ DGs across the five spatial domains identified by BNPSpace. \label{fig:3}}
\end{figure}

The spots of MOB ST data were manually annotated into $K=4$ layers by \cite{ma2022spatially} based on histology and serve as the ground truth. As depicted in Figure \ref{fig:3}(a), BNPSpace achieved the highest level of concordance with the manual annotation and the highest ARI of $0.665$. Using a PPI threshold of $c=0.5$ (see Figure S2(b)), BNPSpace identified $444$ DGs. The gene expression patterns of the top $16$ DGs determined by their PPIs are visualized in Figure S6 of the supplement. We then analyzed differential gene expression (DGE) for those DGs across $\hat{K}=5$ spatial domains. The results illustrated in Figure \ref{fig:3}(c) suggest that each spatial domain is characterized by its unique set of highly expressed genes, aligning with our initial insights discussed in the motivating example in Section \ref{sec:data_motivate}.

To further validate that BNPSpace-defined DGs aligned well with established biological knowledge, we compared our DGs, as well as those SVGs detected by SPARK and SpatialDE \citep{svensson2018spatialde}, another well-known SVG detection method, with known olfactory-bulb-specific genes defined in the Harmonizome database (PMID: 27374120). The DGs, SPARK-identified SVGs, and SpatialDE-identified SVGs are $444$, $572$ and $2,109$. The comparison revealed that the DGs detected by BNPSpace had a higher overlap, odd ratio $=2.223$ ($p$-value $< 0.0001$), with the olfactory-bulb-specific gene set than the SVGs identified by either SPARK or SpatialDE (see Figure \ref{fig:3}(b)). This indicates BNPSpace's effectiveness in discerning biologically relevant gene sets from SRT data. 


\subsection{Application to the Human DLPFC 10X Visium Data}
\label{DLPFC}

\cite{maynard2021transcriptome} utilized the 10x Visium assay to generate spatial molecular profiles for $12$ DLPFC samples. They manually annotated six cortical layers and white matter based on cytoarchitecture, resulting in $K=7$ `true' spatial domains. The data is accessible through the \verb|R| package \verb|spatialLIBD| and is extensively benchmarked in SRT methodological research. Our analysis focused on sample 151509, which comprised $4,788$ spots and $33,538$ genes. Following a quality control procedure similar to that previously described, we obtained preprocessed SRT data of $n=4,786$ spots and $p=1,851$ genes. For the MRF hyperparameter, we selected $d=1.5$ informed by the pBIC plot shown in Figure S7(a) of the supplement. We initiated three independent MCMC chains and performed convergence diagnostics, with the results summarized in Figure S8 of suppelement suggesting good mixing. For the alternative approaches, we used their default settings. The computational efficacy is detailed in Table S4 of the supplement.

Using a PPI threshold of $c=0.5$ (see Figure S7(b)), BNPSpace identified $887$ DGs. The gene expression patterns of the top $16$ DGs determined by their PPIs are visualized in Figure S9 of the supplement. Figure \ref{fig:4}(d) presents the $13$ spatial domains identified by BNPSpace, achieving a moderate ARI of $0.386$. This performance slightly exceeded that of BayesSpace and SpaGCN (ARI $=0.334$ and $0.328$, respectively) when pre-specifying $K=7$ to match the seven manually annotated spatial domains. To refine the spatial clustering results, we applied hierarchical clustering to the $13$ BNPSpace-identified domains based on their pairwise distances, defined as $\text{dist}(k,k')=\sqrt{\sum_{\{j:\hat{\gamma}_j=1\}}(\hat{\mu}_{kj}^*-\hat{\mu}_{k'j}^*)^2}$. The approach enabled us to trim the resulting binary tree (i.e., dendogram) at a specific depth to a more manageable number of clusters $K$. In Figure \ref{fig:4}(b), the red line links the ARIs when pruning the dendrogram into $K=2,\ldots,13$ clusters, while those points in other colors denote the ARIs obtained by other methods for their respective $K$ values. This extensive comparison reveals that BNPSpace consistently surpassed other approaches across various choices of $K$. Notably, BNPSpace reached its highest ARI of $0.553$ at $K=4$, with the corresponding spatial domain pattern present in Figure \ref{fig:4}(a). To understand why BNPSpace identified a larger number of spatial domains, we compared the spatial domain pattern with the expression patterns of six representative DGs, as shown in Figure \ref{fig:4}(e)). Interestingly, genes like \textit{SCGB2A2} and \textit{SCGB1D2} exhibited high expression in the bottom-right area, leading to the division of spatial domains 2, 3, and 4 into subdomains, while the expression patterns of the remaining four DGs closely aligned with the ground truth. We also reported DGE analysis for all $887$ DGs across the $\hat{K}=13$ spatial domains in Figure \ref{fig:4}(c), indicating that each spatial domain was distinguished by its distinct set of highly expressed genes.

To validate that BNPSpace-defined DGs were in alignment with established biological knowledge, we conducted a comparison against the known DLPFC-specific gene list in the Allen Human Brain Atlas (PMID: 23193282). The DGs, SPARK-identified SVGs, and SpatialDE-identified SVGs are $887$, $4,812$ and $3,296$. This comparison revealed that the DGs detected by BNPSpace had a more substantial overlap, odd ratio $=1.877$ ($p$-value $=0.00162$), with the known DLPFC gene set than the SVGs identified by either SPARK or SpatialDE (see Figure \ref{fig:4}(f)). This result confirms that BNPSpace's identified DGs are indeed consistent with well-established biological knowledge.

\section{Conclusion}
\label{sec:conc}

The evident correlation between manually annotated spatial domain patterns and SVGs' expression patterns measured by NSG-based SRT platforms inspired the development of BNPSpace, a novel BNP spatial clustering framework. BNPSpace presents several advantages over existing state-of-the-art methods. First, it directly models count data using an ZIP distribution, offering a more accurate representation of the zero-inflated SRT data. Second, it embeds a feature selection mechanism that not only generates low-dimensional summaries of the SRT data, but also pinpoints the most discriminating features (i.e., DGs). This enhances both the model performance and its interpretability. Most importantly, BNPSpace builds upon an MRF-constraint MFM model to automatically determine the number of clusters (i.e., spatial domains) while encouraging a smoother partitioning. For efficient inference of model parameters, we developed Markov Chain Monte Carlo (MCMC) algorithms based on the Gibbs sampler. In our simulation study, BNPSpace demonstrated superior performance compared to other clustering methods, particularly in scenarios with a high prevalence of zeros in the data. When applied to two real SRT datasets, BNPSpace not only outperformed existing clustering methods, but also uncovered previously unrecognized subdomain structures. The DGs identified by BNPSpace were differentially expressed and linked to significant biological functions.


Our model presents several limitations that merit further investigation. One key limitation is the significant computational demand of the collapsed Gibbs sampler, particularly when dealing with large sample sizes. The time complexity of our algorithm is approximate $\mathcal{O}(npU(K_{max} +1))$. A possible solution to mitigate this issue could be the adoption of a split-merge sampler procedure \citep{jain2004split}. Additionally, it would be interesting to model the extra zero proportion $\pi_i$'s \textit{via} a function related to the latent gene expression levels. This \textit{nonignorable} missing mechanism has already been used in scRNA-seq analysis \citep{wu2022estimating, song2020flexible}.  Furthermore, tuning of~$d$ is criterion-based. A tuning-free algorithm such as putting prior on~$d$ may
improve efficiency. Besides the spatial information, other auxiliary
information, such as cell types, could also be taken into account
for clustering in our future work.

\clearpage

\begin{figure}
\begin{center}
\includegraphics[width = 1\textwidth]{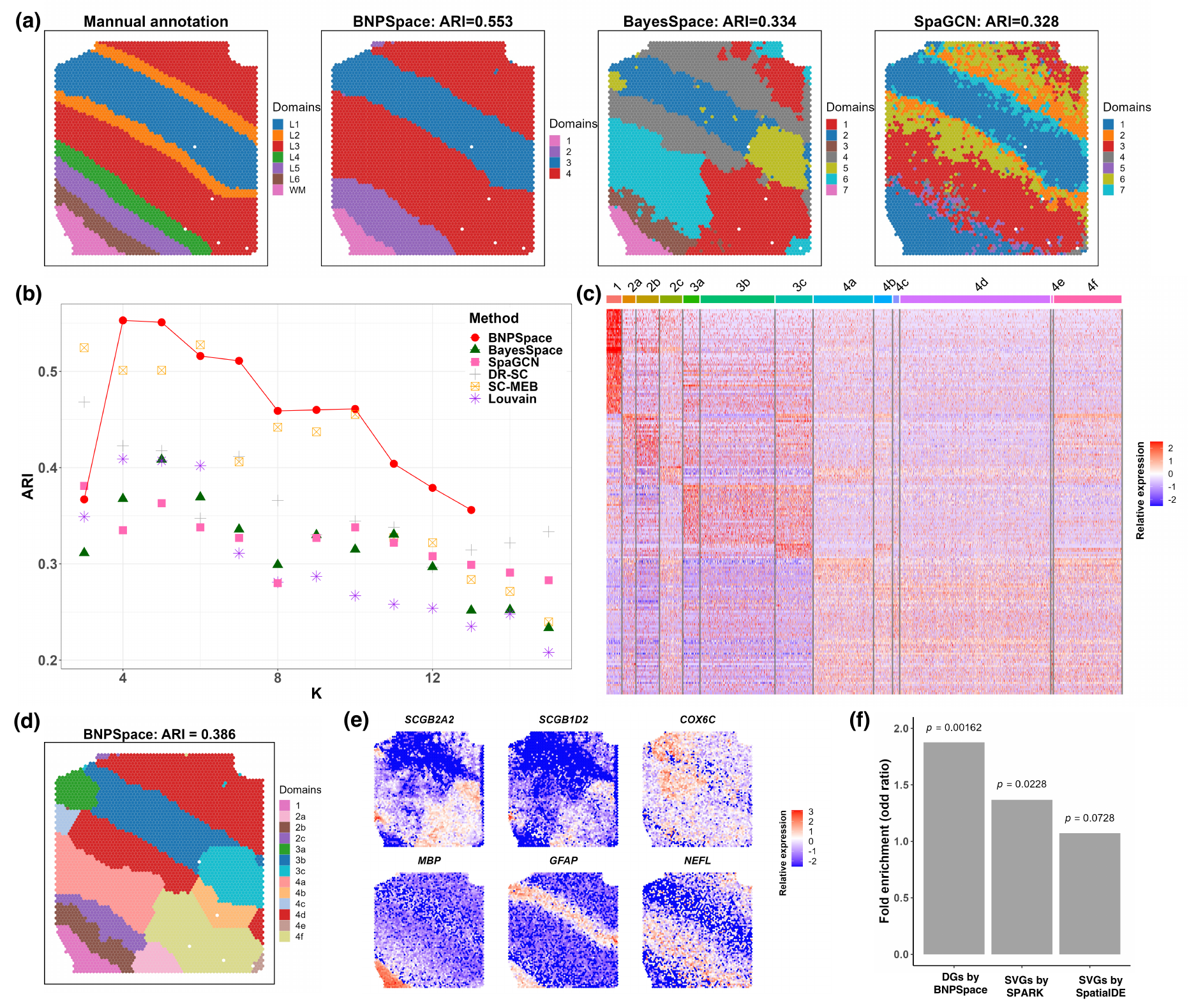}
\end{center}
\caption{Human DLPFC 10x Visium data analysis: (a) Spatial domains annotated by pathologists and identified by BNPSpace (after postprocessing) and competing methods; (b) ARIs achieved by BNPSpace and competing methods across various $K$; (c) Heatmap of $887$ DGs across the $13$ spatial domains identified by BNPSpace; (d) $13$ spatial domains identified by BNPSpace; (e) Spatial expression patterns of six representative DGs; (f) Gene enrichment analysis for $887$ BNPSpace-identified DGs, $4812$ SPARK-identified SVGs, and $3296$ SpatialDE-identified SVGs.  \label{fig:4}}
\end{figure}

\clearpage
\renewcommand{\thetable}{S\arabic{table}}%
\renewcommand{\thefigure}{S\arabic{figure}}%
\renewcommand{\thesection}{S\arabic{section}}%
\setcounter{section}{0}
\renewcommand{\theequation}{S\arabic{equation}}%

\section{P\'{o}lya Urn Scheme of MRF-MFM model}

\begin{theorem}
Under the following Markov random field constrained mixture of finite mixtures (MRF-constrained MFM) model specification,
$$
\begin{aligned}
P(\bm{z} \mid  \bm{q}) \propto\left(\prod_{i=1}^n q_{z_i}\right) \exp \left(d \sum_{i < i'}g_{ii'} {I}(z_i=z_{i'}) \right), \notag \\
    (q_{1}, \ldots, q_{K}) | K \sim \operatorname{Dir}(\alpha_{0}, \ldots, \alpha_{0}), \quad K-1 \sim P_{\theta}(K).
\end{aligned}
$$
The P\'{o}lya Urn Scheme is
$$
\Pi\left(z_i=k \mid \bm{z}_{-i}\right) \propto\left\{\begin{array}{ll}
{[n_{k, -i} + \alpha_{0}] \exp \left[d \sum_{i' = 1}^{n} g_{ii'} I\left(z_{i'} = k\right)\right],} & \text { at an existing cluster } k \\
V_n\left(|\bm{z}_{-i}|+1\right) / V_n\left(|\bm{z}_{-i}|\right) \alpha_{0}, & \text { if } k \text { is a new cluster }
\end{array},\right.
$$
where $\bm{z}_{-i}=\bm{z} \backslash\left\{z_i\right\}$, i.e., all elements of $\bm{z}$ except for $z_i$, $|\bm{z}_{-i}|$ is the number of clusters in $\bm{z}_{-i}$, and $V_{n}(t) = \sum_{K=1}^{\infty} \frac{\tbinom{K}{t} t!}{(K\alpha_{0})^{(n)}} P_{\theta}(K) = \sum_{K=1}^{\infty} \frac{t! \tbinom{K}{t} \Gamma(K\alpha_{0})}{ \Gamma(n + K\alpha_{0})} P_{\theta}(K)$. \label{thm1}
\end{theorem}
~\\

\begin{proof}
Integrating out mixing parameter $\bm{q}$, 
$$
\begin{aligned}
P(\bm{z} \mid K) & \propto \int P(\bm{z} \mid \bm{q}, K) P(\bm{q} \mid K)d(\bm{q}) \\
& \propto \int \prod_{i=1}^{n} q_{z_{i}} \exp\left(d \sum_{i < i' } g_{ii'} {I}(z_{i} = z_{i'})\right) \frac{\Gamma(K\alpha_{0})}{\Gamma(\alpha_{0})^{K}} q_{1}^{\alpha_{0} -1} \cdots q_{K}^{\alpha_{0} -1} d\bm{q} \\
& \propto \frac{\Gamma(K\alpha_{0})}{\Gamma(\alpha_{0})^{K}} \frac{\prod_{l=1}^{K} \Gamma(n_{l} + \alpha_{0})}{\Gamma(n + K \alpha_{0})} \exp\left(d \cdot \sum_{i< i' } g_{ii'} {I}(z_{i} = z_{i'})\right),
\end{aligned}
$$
where $n_{l} = |c_{l} = \{i: z_{i} = l\}|$. Let $\mathcal{C} = \{c_{1}, \ldots , c_{t}\}$ be the $t$ partitions induced by $\bm{z} = (z_{1}, \ldots, z_{n})^{\top}$, then
$$
P\left(\mathcal{C}, \bm{z} \mid K\right) \propto \frac{1}{(K\alpha_{0})^{(n)}} \prod_{c \in \mathcal{C}}\alpha_{0}^{(|c|)} \exp(d \cdot  E_{c} ),
$$
where we define $m^{(n)} = \frac{\Gamma(m + n)}{\Gamma(m)}$ and $E_{c} = \sum_{i, i' \in c}g_{ii'}$. Hence, 
$$
P(\mathcal{C} \mid K) \propto \sum_{\bm{z}: \mathcal{C}(\bm{z}) = \mathcal{C}} P\left(\mathcal{C}(\bm{z}), \bm{z} \mid K\right) \propto \frac{\tbinom{K}{t} t!}{(K\alpha_{0})^{(n)}} \prod_{c \in \mathcal{C}}\alpha_{0}^{(|c|)} \exp(d \cdot E_{c} ).
$$
Then, 
$$
P(\mathcal{C}) \propto \sum_{K=1}^{\infty} P(\mathcal{C} \mid K)P_{\theta}(K) \propto \left(\sum_{K=1}^{\infty} \frac{\tbinom{K}{t} t!}{(K\alpha_{0})^{(n)}} P_{\theta}(K)\right) \prod_{c \in \mathcal{C}}\alpha_{0}^{(|c|)} \exp(d \cdot E_{c}) 
\propto V_{n}(t) \prod_{c \in \mathcal{C}}\alpha_{0}^{(|c|)} \exp(d \cdot E_{c}),
$$
where $V_{n}(t) = \sum_{K=1}^{\infty} \frac{\tbinom{K}{t} t!}{(K\alpha_{0})^{(n)}} P_{\theta}(K)$.

We denote $\mathcal{C}_{n}$ is the partition induced by $n$ samples. Then we can derive the conditional distribution of $\mathcal{C}_{n} | \mathcal{C}_{n-1}$. If $\{n\}$ is a singleton  in $\mathcal{C}_{n}$, then
$$
P(\mathcal{C}_{n} \mid \mathcal{C}_{n-1}) \propto  V_{n}(t +1) \prod_{c \in \mathcal{C}_{n-1}}\alpha_{0}^{(|c|)} \exp(d \cdot E_{c})  \alpha_{0}^{(1)},
$$
if $\{n\}$ belongs to a existing cluster $l$, then
$$
P(\mathcal{C}_{n} \mid \mathcal{C}_{n-1}) \propto  V_{n}(t) \prod_{c \in \mathcal{C}_{n-1}\setminus c_{l}}\alpha_{0}^{(|c|)} \exp(d \cdot E_{c} ) \alpha_{0}^{(|c_{l}| + 1)} \exp(d \cdot E_{c_{l}} + d \cdot E_{c_l}^{n}), 
$$
where $E_{c_l}^{n} = \sum_{i=1}^{n-1}g_{ni}I(z_{i} = l)$. Hence,
$$
P(\mathcal{C}_{n} \mid \mathcal{C}_{n-1}) \propto 
\begin{cases} (\alpha_{0}+|c|)\exp(d \cdot E_{c}^{n}) & \text { if } c \in \mathcal{C}_{n-1} \text { and } c \cup\{n\} \in \mathcal{C}_n \\
V_n(t+1)/ V_n(t) \alpha_{0} & \text { if } n \text { is a singleton in } \mathcal{C}_n \text {, i.e., }\{n\} \in \mathcal{C}_n \\
\end{cases}.
$$
\end{proof}

\section{Full details of MCMC procedure}
The molecular profile data is denoted by matrix notation $\bm{Y}_{n \times p}$. Parameter is denoted by $\Theta = (\bm{R}_{n \times p}, \bm{\gamma}, \bm{z}, \bm{M}_{n\times p})$ where $\bm{R}$ are extra zero indicators, $\bm{z}$ are cluster allocation indicators, $\gamma$ are the DG indicators, and $\bm{M}$ are the parameters denoting normalized latent gene expression. The data likelihood is given by
$$
L(\bm{Y} | \bm{R}, \bm{M}, \bm{\gamma}, \bm{z}) = \prod_{i=1}^{n} \prod_{j = 1}^{p} P(y_{ij} \mid r_{ij}, \gamma_{j}, z_{i}, \mu_{ij}).
$$
In detail, the probability
$$P(y_{ij} \mid r_{ij}, \gamma_{j}, z_{i} = k, \mu_{ij}) = \begin{cases}
   {\begin{array}{ll}
		 I(y_{ij} = 0) & \text{ if }  r_{ij}=1 \\
			\text{Poi}(s_i\mu_{kj}^{*}) & \text{ if } r_{ij}=0 \text{ and } \gamma_j=1 \\
		\text{Poi}(s_i\mu_{0j}) & \text{ if } r_{ij}=0 \text{ and } \gamma_j=0
	\end{array}}
\end{cases}$$
is the likelihood of the following model, 
$$
y_{ij} | r_{ij}, \mu_{kj}^*,\mu_{0j},\gamma_j,z_i=k\sim
	\begin{cases}
		r_{ij}I(y_{ij}=0)+(1-r_{ij})\text{Poi}\left(s_{i}\mu_{kj}^*\right)& \text{ if } \gamma_j=1\\
		r_{ij}I(y_{ij}=0)+(1-r_{ij})\text{Poi}\left(s_{i}\mu_{0j}\right) & \text{ if } \gamma_j=0
        \end{cases},
$$
where $\gamma_{j}$ is a DG indicator to imply that the expression of gene $j$ is varied among different clusters. 

The priors of parameters are given by
\begin{equation}
     \gamma_{j} \sim \operatorname{Bern}(\omega) \quad  \omega \sim \operatorname{Be}(\alpha_{\omega}, \beta_{\omega}), \notag
\end{equation}
\begin{equation}
    r_{ij} \sim \operatorname{Bern}(\pi_{i}) \quad \quad \pi_{i} \sim \operatorname{Be}(\alpha_{\pi}, \beta_{\pi}),  \notag
\end{equation}
\begin{equation}
    \mu_{kj}^{*} \sim \operatorname{Ga}(\alpha_{\mu}, \beta_{\mu}) \quad \quad \mu_{0j} \sim \operatorname{Ga}(\alpha_{\mu}, \beta_{\mu}), \notag
\end{equation}
for $i = 1, \dots, n$, $j =1, \ldots, p$ , and $k = 1, \ldots K$. 

A Markov random field constrained mixture of finite mixtures (MRF-constrained MFM) prior is given for the cluster allocation $\bm{z}$, 
$$
\begin{aligned}
P(\bm{z} \mid \bm{q}) \propto\left(\prod_{i=1}^n q_{z_i}\right) \exp \left(d \sum_{i < i'} g_{ii'} {I}(z_i=z_{i'}) \right), \quad 
\bm{q} | K \sim \operatorname{Dir}(\bm{\alpha}_{0}), \quad K-1 \sim \operatorname{Poi}(\lambda).
\end{aligned}
$$
The constraint implies that the neighboring spots are more likely to belong to the same cluster. Hence, the posterior of parameters is given by
$$
P(\bm{R}, \bm{\gamma}, \bm{z}, \bm{M} \mid \bm{Y}) \propto P(\bm{Y} | \bm{R}, \bm{\gamma}, \bm{z}, \bm{M}) P(\bm{R}) P(\bm{M} | \bm{\gamma}, \bm{z}) P(\bm{\gamma}) P(\bm{z}).
$$
~\\ 
\textbf{Step 1: Update the DG indicator vector} $\bm{\gamma}$ \\
The posterior conditional distribution of $\bm{\gamma}$ is given by
$$
P(\bm{\gamma} \mid \cdot)  \propto P(\bm{Y} | \bm{R}, \bm{\gamma},  \bm{z}, \bm{M}) P(\bm{M} \mid \bm{\gamma}, \bm{z}) P(\bm{\gamma} \mid \alpha_{\omega}, \beta_{\omega}),
$$
where $P(\bm{\gamma}\mid \alpha_{\omega}, \beta_{\omega}) = \int \prod_{j=1}^{p} P(\gamma_{j} | \omega)P(\omega \mid \alpha_{\omega}, \beta_{\omega})d\omega \propto \Gamma(\alpha_{\omega} +  \sum_{j= 1}^{p}\gamma_{j})\Gamma(\beta_{\omega} + p - \sum_{j= 1}^{p}\gamma_{j})$.
By collapsed Gibbs sampler, We can integrate out parameters $\mu_{kj}^{*}$ and $\mu_{0j}$ given $\bm{\gamma}$ and $\bm{z}$ for $j = 1, \ldots, p$ and $k = 1, \ldots, K$, obtaining
$$
\begin{aligned}
& P(\bm{Y} | \bm{R}, \bm{\gamma}, \bm{z}) \\
= & \prod_{j=1}^{p} \prod_{k=1}^{K} \int \prod_{\{i:z_{i} = k\}} P({y}_{ij} \mid {r}_{ij}, {\gamma_{j}}, z_{i} = k, \mu_{kj}^{*}, \mu_{0j})P(\mu_{kj}^{*}) P(\mu_{0j}) d\mu_{kj}^{*} d\mu_{0j} \\
= & \prod_{\{j:\gamma_{j}=1\}} \prod_{k=1}^{K} \int \prod_{\{i:z_{i} = k, r_{ij} = 0\}}   \operatorname{Poi}(y_{ij}|s_{i} \mu_{kj}^{*})\operatorname{Ga}(\mu_{kj}^{*}\mid \alpha_{\mu}, \beta_{\mu})d\mu_{kj}^{*} \\
& \quad \quad\times \prod_{\{j:\gamma_{j}=0 \}} \int \prod_{\{i:r_{ij} = 0\}} \operatorname{Poi}(y_{ij}|s_{i} \mu_{0j}) \operatorname{Ga}(\mu_{0j}\mid \alpha_{\mu}, \beta_{\mu})d\mu_{0j} \\
= & \left(\prod_{j=1}^{p} \prod_{\{i:r_{ij} = 0\}} \frac{s_{i}^{y_{ij}}}{y_{ij}!} \right) 
\left(\prod_{\{j:\gamma_{j}=1\}} \prod_{k=1}^{K} \frac{\beta_{\mu}^{\alpha_{\mu}}}{\Gamma(\alpha_{\mu})} \frac{\Gamma(\alpha_{\mu} + \sum_{\{i:z_{i} = k, r_{ij} = 0\}} y_{ij})}{(\beta_{\mu} + s_{i})^{\alpha_{\mu} + \sum_{\{i:z_{i} = k, r_{ij} = 0\}} y_{ij}}}\right)  \\
& \quad \quad \times \left(\prod_{\{j:\gamma_{j}=0\}} \frac{\beta_{\mu}^{\alpha_{\mu}}}{\Gamma(\alpha_{\mu})} \frac{\Gamma(\alpha_{\mu} + \sum_{\{i: r_{ij} = 0\}} y_{ij})}{(\beta_{\mu} + s_{i})^{\alpha_{\mu} + \sum_{\{i: r_{ij} = 0\}}y_{ij}}}\right).
\end{aligned}
$$
As a result, the marginalized conditional posterior of $\bm{\gamma}$ is 
$$
P(\bm{\gamma} \mid \cdot) \propto P(\bm{Y} | \bm{R}, \bm{\gamma}, \bm{z}) P(\bm{\gamma} \mid \alpha_{\omega}, \beta_{\omega}),
$$

We adopt a stochastic search procedure to sample indicators $\bm{\gamma}$. Given the indicators $\bm{\gamma}^{\text{old}}$, we choose a move between Add/Delete move and Swap with probability $\rho$. The Add/Delete move randomly selects one of the elements in $\bm{\gamma}^{\text {old}}$ and changes its value (from $0$ to $1$ or from $1$ to $0$ ). The Swap move randomly selects two elements in $\bm{\gamma}^{\text{old}}$ with different inclusion status and swaps their values. A new candidate $\bm{\gamma}^{\text{new}}$ is accepted with probability
$$
\min \left\{\frac{P\left(\bm{Y} \mid \bm{R}, \bm{\gamma}^{\text {new }}, \bm{z}\right) P\left(\bm{\gamma}^{\text {new }} \mid \alpha_{\omega}, \beta_{\omega} \right)} {P\left(\bm{Y} \mid \bm{R}, \bm{\gamma}^{\text {old }}, \bm{z}\right) P\left(\bm{\gamma}^{\text {old }} \mid \alpha_{\omega}, \beta_{\omega} \right)}, 1 \right\}.
$$

~\\
\textbf{Step 2: Update the spatial domain allocation vector} $\bm{z}$  \\
By Theorem \ref{thm1}, we implement the Gibbs sampling algorithm to update $z_{i}$ sequentially from $i = 1$ to $n$ at each iteration. 

The probability of $z_{i}$ belonging to an existing cluster is
$$
\begin{aligned}
& \quad P(z_{i} = k\mid \bm{R}, \bm{z}_{-i}, \bm{\gamma}, \bm{M}, \bm{Y}) \\
& \propto (n_{k, -i} + \alpha_{0}) \exp [d \sum_{i' = 1}^{n} g_{ii'} I\left(z_{i'} = k\right)] P(\bm{y}_{i \cdot}\mid \bm{r}_{i\cdot}, \bm{\gamma}, \bm{M}) \\
& \propto (n_{k, -i} + \alpha_{0}) \exp [ d \sum_{i' = 1}^{n} g_{ii'} I\left(z_{i'} = k\right)] \prod_{\{j: \gamma_{j}=1, r_{ij}=0\}} \operatorname{Poi}(y_{ij}|s_{i}\mu_{kj}^{*})
\end{aligned}
$$
where $n_{k, -i}$ is the number of spots belonging to cluster $k$ after removing spot $i$. 

The probability of $z_{i}$ belonging to a new cluster is 
$$
\begin{aligned}
& \quad P(z_{i} = K+1 \mid \bm{r}_{i\cdot}, \bm{\gamma}, \bm{y}_{i\cdot}) \\
& \propto \alpha_{0} \frac{V_n(K+1)}{V_n(K)} \prod_{\{j: \gamma_{j} = 1\}} P({y}_{ij} \mid {r}_{ij}, s_{i}, \gamma_{j} = 1) \\
& \propto \alpha_{0} \frac{V_n(K+1)}{V_n(K)} \int \prod_{\{j: \gamma_{j} = 1\}}  P(y_{ij}|\bm{r}_{i}, s_{i}, \gamma_{j} = 1, \mu_{ij}) H(d\mu_{ij})\\
& \propto \alpha_{0} \frac{V_n(K+1)}{V_n(K)} \prod_{\{j:\gamma_{j}=1, r_{ij}= 0\}} \frac{s_{i}^{y_{ij}}}{y_{ij}!} \frac{\beta_{\mu}^{\alpha_{\mu}}}{\Gamma(\alpha_{\mu})} \frac{\Gamma(\alpha_{\mu} + y_{ij})}{(\beta_{\mu} + s_{i})^{\alpha_{\mu} + y_{ij}}} ,
\end{aligned}
$$
where the coefficient $V_{n}(K)$ could be calculated before MCMC procedure. 

Then, we normalize $P(z_{i} = k | \cdot)$ for $k=1, \ldots, K+1$, obtaining the probability $p_{k} = \frac{P(z_{i} = k | \cdot)}{\sum_{m=1}^{K+1}P(z_{i} = m | \cdot)}$, and sample $z_{i}$ from following categorical distribution with probability $\bm{p}^{*} = (p_{1}, \ldots, p_{K+1})^{\top}$ by $z_{i} \sim  \operatorname{Mult}(1, \bm{p}^{*})$.  

~\\
\textbf{Step 3: Update the latent normalized gene expression parameter} $\bm{M}$ \\
If $\gamma_{j} = 1$, update $\mu_{kj}^{*}$ by 
$$
\mu_{kj}^{*} | \cdot \sim \operatorname{Ga}(\alpha_{\mu} + \sum_{\{i:z_{i} = k, r_{ij} = 0\} } y_{ij}, \quad \beta_{\mu} + \sum_{\{i:z_{i} = k, r_{ij} = 0\} } s_{i}).
$$
If $\gamma_{j} = 0$, update $\mu_{0j}$ by 
$$
\mu_{0j} |\cdot \sim \operatorname{Ga}(\alpha_{\mu} + \sum_{\{i: r_{ij} = 0\} } y_{ij}, \quad \beta_{\mu} + \sum_{\{i: r_{ij} = 0\} } s_{i}).
$$

~\\
\textbf{Step 4: Update the extra zero indicator} $\bm{R}$\\
If $\gamma_{j} = 1$, we update $r_{ij} \mid y_{ij} = 0, z_{i} =k$ by 
$$
r_{ij} | \cdot \sim \operatorname{Bern}(\frac{\pi_{i}}{\pi_{i} + (1-\pi_{i})\exp(-s_{i}\mu_{kj}^{*})}).
$$
If $\gamma_{j} = 0$, we update $r_{ij} \mid y_{ij} = 0$ by 
$$
r_{ij} | \cdot \sim \operatorname{Bern}(\frac{\pi_{i}}{\pi_{i} + (1-\pi_{i})\exp(-s_{i}\mu_{0j})}).
$$

~\\
\textbf{Step 5: Update the underlying proportions of extra zero} $\bm{\pi}$ \\
We update $\pi_{i}$ for $i = 1, \ldots, n$ by the Gibbs sampler
$$
\pi_{i} | \cdot \sim \operatorname{Be}(\alpha_{\pi} + A_{i}, \beta_{\pi} + p - A_{i}),
$$
where $A_{i} = \sum_{j}r_{ij}$. 

\section{Selection of hyperparameter $d$ in the MRF prior}
In the MRF-MFM model, choosing the hyperparameter $d$ in the Markov random field is important, which controls the magnitude of spatial smoothness. Higher $d$ leads to higher spatial smoothness, resulting in a smaller number of clusters and decreasing the model complexity. We recommended using the modified Bayesian information criteria (penalized BIC) \citep{pan2007penalized} to find an appropriate $d$. The pBIC is calculated as
\begin{equation}
    \text{pBIC}(d) = -2 \log(L(\bm{Y} \mid \hat{\bm{M}}, \hat{\bm{R}}, \hat{\bm{z}}, \hat{\bm{\gamma}}, d )) + \log(n)(\hat{p}_{\gamma}*\hat{K} + p - \hat{p}_{\gamma}), \nonumber
\end{equation}
where $\hat{p}_{\gamma}$ is the estimated number of discriminating genes and $\hat{K}$ is the estimated number of clusters,   $\hat{\mu}_{kj}^{*}$ and $\hat{\mu}_{0j}$ for $j =1, \ldots, p$ and $k = 1, \dots, \hat{K}$ are the posterior means of corresponding parameters, and  $\hat{\bm{R}}$ and $\hat{\bm{\gamma}}$ are the estimators based on PPIs with threshold $0.5$. In the formula, 
\begin{equation}
    L(\bm{Y} \mid \hat{\bm{M}}, \hat{\bm{R}}, \hat{\bm{z}}, \hat{\bm{\gamma}}, d ) 
    = \prod_{\{j: \hat{\gamma}_{j} = 1\}} \prod_{k=1}^{\hat{K}} \prod_{\{i:\hat{z}_{i} =k, \hat{r}_{ij} = 0\}} \operatorname{Poi}(y_{ij}|s_{i}\hat{\mu}_{kj}^{*}) \prod_{\{j: \hat{\gamma}_{j} = 0\}}\prod_{\{i:\hat{r}_{ij} = 0\}} \operatorname{Poi}(y_{ij}|s_{i}\hat{\mu}_{0j}).
    \nonumber
\end{equation}

\section{Simulation Study}
We use both simulated spatial patterns and a real spatial pattern to evaluate the performance of BNPSPace. 

\subsection{Data Generation}
\begin{figure}
\renewcommand*\thefigure{S\arabic{figure}}
\begin{center}
\includegraphics[width = 1\textwidth]{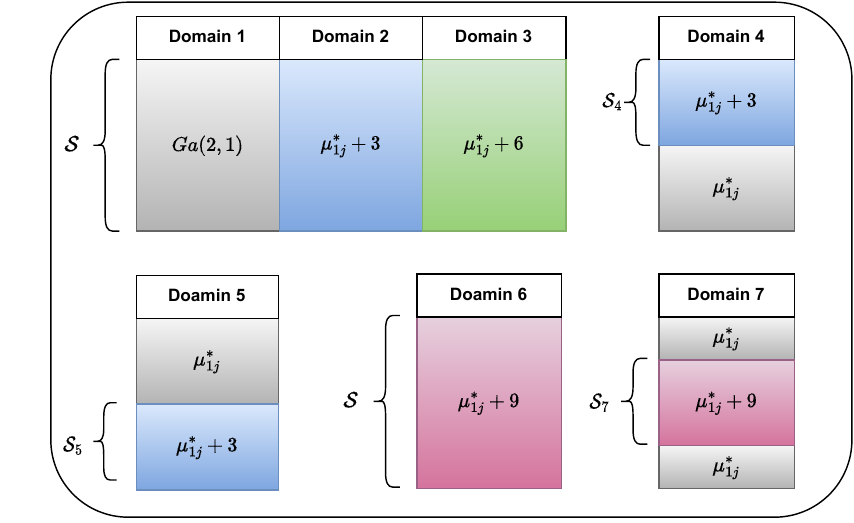}
\end{center}
\caption{Simulated data: An illustration of the process used to generate $\mu_{kj}^{*}$ for simulated data, where $\mathcal{S}$ represents the set of DGs. \label{fig:s2}}
\end{figure}

Let $\bm{Y}_{n \times p}$ denote the simulated count table.  We set the number of DGs $p_{\gamma} = 20$ and total genes $p = 1000$. The size factor was $s_{i} \sim \operatorname{Unif}(0.5, 1.5)$ for $i = 1, \ldots, n$. The nonDGs expression level, $\mu_{0j} \sim \operatorname{Ga}(2,1)$. The latent gene expression level $\mu_{kj}^{*}$ was generated by different schemes corresponding to the number of clusters $K = 3, 5$, and $7$. The pattern of $\mu_{kj}^{*}$ is shown in the following paragraph and Figure \ref{fig:s2}. Finally, the count expression matrix was generated by
$$
y_{ij} | \pi_{i}, \mu_{kj}^*,\mu_{0j},\gamma_j, z_i=k\sim
	\begin{cases}
		\pi_{i}I(y_{ij}=0)+(1-\pi_{i})\text{Poi}\left(s_{i}\mu_{kj}^*\right)& \text{ if } \gamma_j=1\\
		\pi_{i}I(y_{ij}=0)+(1-\pi_{i})\text{Poi}\left(s_{i}\mu_{0j}\right) & \text{ if } \gamma_j=0
	\end{cases},
$$
with low, medium ,and high zero-inflation levels where $\pi_{i} = 0.1, 0.2$ and $0.3$. 

The expression levels of all discriminating genes are different for the first three domains. The latent expression level of domain 1 followed a gamma distribution given by $\mu_{1j}^{*} \sim \operatorname{Ga}(2, 1)$ for $ j = 1, \ldots, p_{\gamma}$, which implied the first $p_{\gamma}$ genes were discriminatory features, denoted by set $\mathcal{S}$. For the second domain and third domain, the latent expression levels were $\mu_{2j}^{*} = \mu_{1j}^{*} + 3$, $\mu_{3j}^{*}  = \mu_{1j}^{*} + 6$ for $j \in \mathcal{S}$.

In the $4$-th domain and $5$-th domain, the discriminating gene sets are different compared to the first domain. The $4$-th domain is generated by the scheme where  $p_{\gamma} /2$ variables (set $\mathcal{S}_{4}$) were randomly selected from the whole discriminatory features ($\mathcal{S}$), then $\mu_{4j}^{*}  = \mu_{1j}^{*} + 3$ for $ j \in \mathcal{S}_{4}$ and $\mu_{4j}^{*} = \mu_{j1}^{*}$ for $j \in \mathcal{S} \setminus \mathcal{S}_{4}$. The expression level $\mu_{5j}^{*}$ of $5$-th cluster was given by $\mu_{5j}^{*}  = \mu_{1j}^{*} + 3$ for $j \in \mathcal{S}_{5} = \mathcal{S}\setminus\mathcal{S}_{4}$ and $\mu_{5j}^{*} = \mu_{1j}^{*}$ for $j \in \mathcal{S}_{4}$. 

In the $6$-th domain and $7$-th domain, the discriminating genes are highly expressed. 
The expression level of $6$-th cluster was $\mu_{6j}^{*}  = \mu_{1j}^{*} + 9$ for $j \in \mathcal{S}$. The discriminating gene set $\mathcal{S}_{7}$ for $7$-th cluster was generated by randomly sampling $p_{\gamma} /2$ features from set $\mathcal{S}$ and we set $\mu_{7j} = \mu_{1j}^{*} + 9$ for $ j \in \mathcal{S}_{7}$ and $\mu_{7j}^{*} = \mu_{1j}^{*}$ for $j \in \mathcal{S}\setminus \mathcal{S}_{7}$. 

\subsection{Prior Specification}
For prior specifications, we use the following setting as default. We set $\alpha_{\mu} = \beta_{\mu} = 1$ to be weakly informative and $\alpha_{\pi}= \beta_{\pi} = 1$ to be a uniform prior, respectively. We set $\alpha_{\omega} = 0.1$ and $\beta_{\omega} = 1.9$ to make the proportion of discriminating features $5 \%$. For the hyperparameters in the MRF-constrained MFM model, we set $\alpha_{0} =1$, $\lambda = 1$ as default, and selected $d$ according to the pBIC criteria. We ran the 10000 MCMC iterations with a burn-in of 5000 iterations. 

\subsection{Evaluation metrics}
We evaluated the performance of BNPSpace in several aspects: clustering performance, identification of the number of clusters, and feature selection performance. For clustering results, we quantified performance via the Adjusted Rand Index (ARI) \citep{hubert1985comparing}, a variant of Rand Index (RI), based on estimation of cluster assignment $\bm{\hat{z}} = (\hat{z}_{1}, \ldots, \hat{z}_{n})^{\top}$. Let $A = \sum_{i > i^{'}}{I}(z_{i} = z_{i^{'}}){I}(\hat{z}_{i} = \hat{z}_{i^{'}})$ be the number of pairs belonging to the same group in the truth and estimation; $B = \sum_{i > i^{'}}{I}(z_{i} = z_{i^{'}}){I}(\hat{z}_{i} \neq \hat{z}_{i^{'}})$ be number of pairs which belongs to the same group in the truth but different groups in the estimation; $C = \sum_{i > i^{'}}{I}(z_{i} \neq z_{i^{'}}){I}(\hat{z}_{i} = \hat{z}_{i^{'}})$ be the number of pairs belonging to different groups in the true partition but assigned to the same group in the estimation; $D = \sum_{i > i^{'}}{I}(z_{i} \neq z_{i^{'}}){I}(\hat{z}_{i} \neq \hat{z}_{i^{'}})$, the number of pairs assigned to different groups in both truth and estimation. Then the ARI is defined as
\begin{equation}
    \text{ARI}=\frac{\left(\begin{array}{l}
n \\
2
\end{array}\right)(A+D)-[(A+B)(A+C)+(C+D)(B+D)]}{\left(\begin{array}{l}
n \\
2
\end{array}\right)^2-[(A+B)(A+C)+(C+D)(B+D)]} .\nonumber
\end{equation}
The larger value of ARI indicates a more accurate clustering result. 

Since BNPSpace also enables the selection of discriminating genes, we quantified its performance via sensitivity, specificity, Matthew’s correlation coefficient (MCC), and area under the curve (AUC) of the receiver operating characteristic based on the PPIs of all genes. Given the threshold of PPIs, the metrics  sensitivity, specificity and MCC can be calculated based on confusing matrix:

\begin{equation}
    \begin{aligned}
       \text{Sensitivity} & = \frac{\text{TP}}{\text{TP} + \text{FN}}, \\
       \text{Specificity} & = \frac{\text{TN}}{\text{TN} + \text{FP}}, \\
       \text{MCC} & = \frac{\text{TP} \times \text{TN} - \text{FP} \times \text{FN}}{\sqrt{(\text{TP} + \text{FP})(\text{TP} + \text{FN}) (\text{TN} + \text{FP}) (\text{TN} + \text{FN})}} .
\end{aligned} \nonumber
\end{equation}
with TP, TN, FP, and FN denoting the true positives, true negatives, false positives, and false negatives, respectively. All these metrics are between $0$ and $1$ and a higher value illustrates a better performance.

\begin{figure}
\renewcommand*\thefigure{S\arabic{figure}}
\begin{center}
\includegraphics[width = 1.1\textwidth]{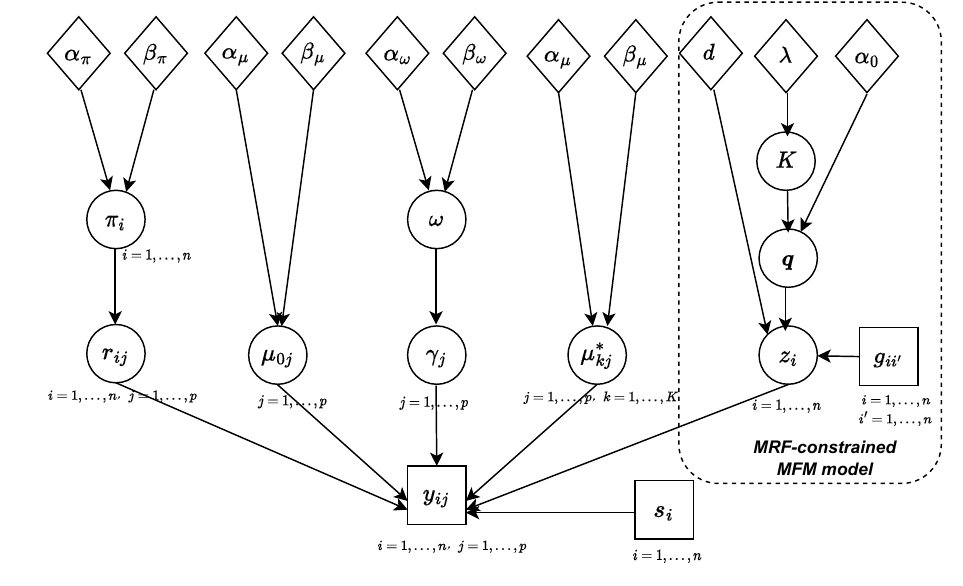}
\end{center}
\caption{Graphical representation of the proposed BNPSpace model. Observable data, model parameters, and fixed hyperparameters are denoted by square, circle, and diamond-shaped nodes, respectively. A link between two nodes represents a direct probabilistic dependence. \label{fig:s1}}
\end{figure}

\begin{figure}
\renewcommand*\thefigure{S\arabic{figure}}
\begin{center}
\includegraphics[width = 1\textwidth]{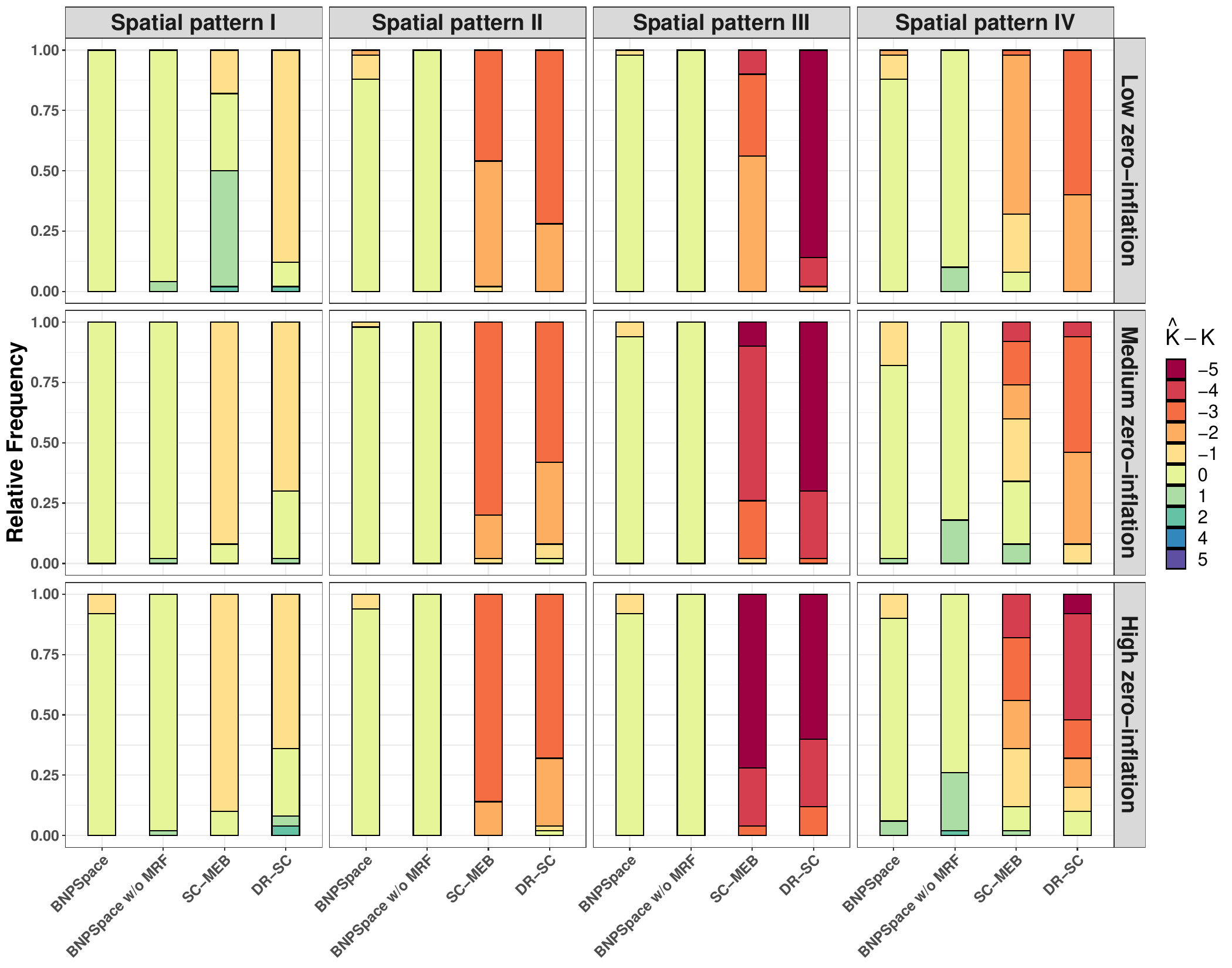}
\end{center}
\caption{Simulated data analysis: Percent stacked barplots of the discrepancies between the estimated number of clusters ($\hat{K}$) and their true values ($K$) achieved by BNPSpace and competing methods across various scenarios in terms of spatial domain patterns and zero-inflation settings.\label{fig:s3}}
\end{figure}

\begin{figure}
\renewcommand*\thefigure{S\arabic{figure}}
\begin{center}
\includegraphics[width = 1\textwidth]{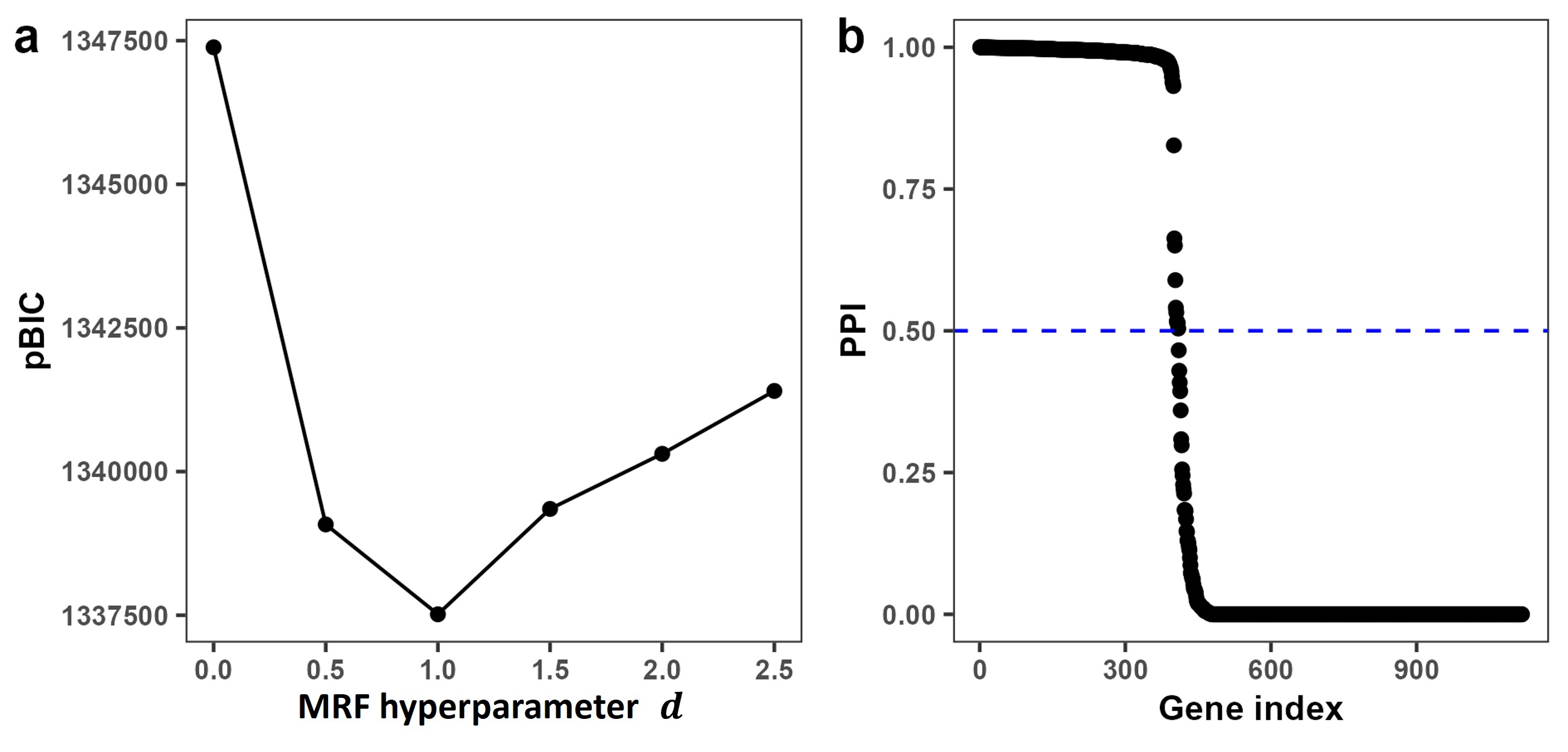}
\end{center}
\caption{MOB ST data analysis: (a) The pBIC across various values of $d$, aiding in the selection of this MRF hyperparameter; (b) The sorted PPI values for $p=1,117$ genes by BNPSpace, aiding in the selection of DGs.\label{fig:s4}}
\end{figure}

\begin{figure}
\renewcommand*\thefigure{S\arabic{figure}}
\begin{center}
\includegraphics[width = 1\textwidth]{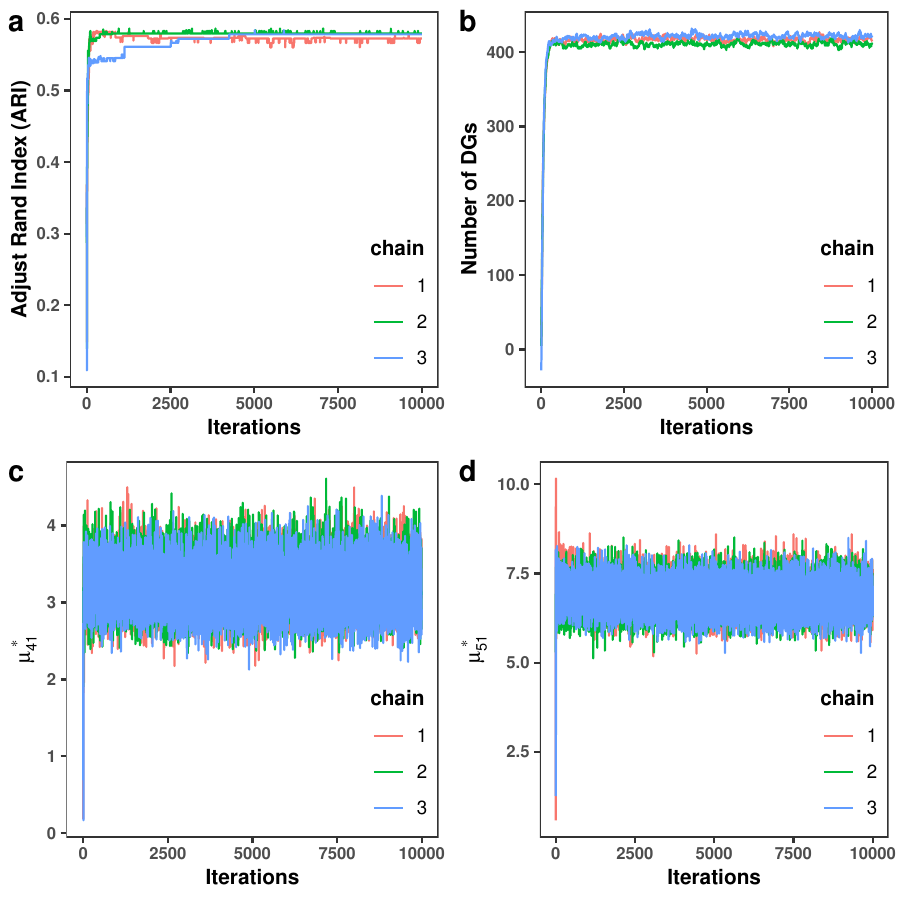}
\end{center}
\caption{MOB ST data analysis: Trace plots of (a) ARI, (b) The number of DGs $p_\gamma$, (c) the normalized expression level $\mu_{41}^*$ for gene $1$ across all spots in spatial domain $4$, and (d) the normalized expression level $\mu_{51}^*$ for gene $1$ across all spots in spatial domain $5$, of three independent MCMC chains with diverse initializations. \label{fig:s5}}
\end{figure}

\begin{figure}
\renewcommand*\thefigure{S\arabic{figure}}
\begin{center}
\includegraphics[width = 1\textwidth]{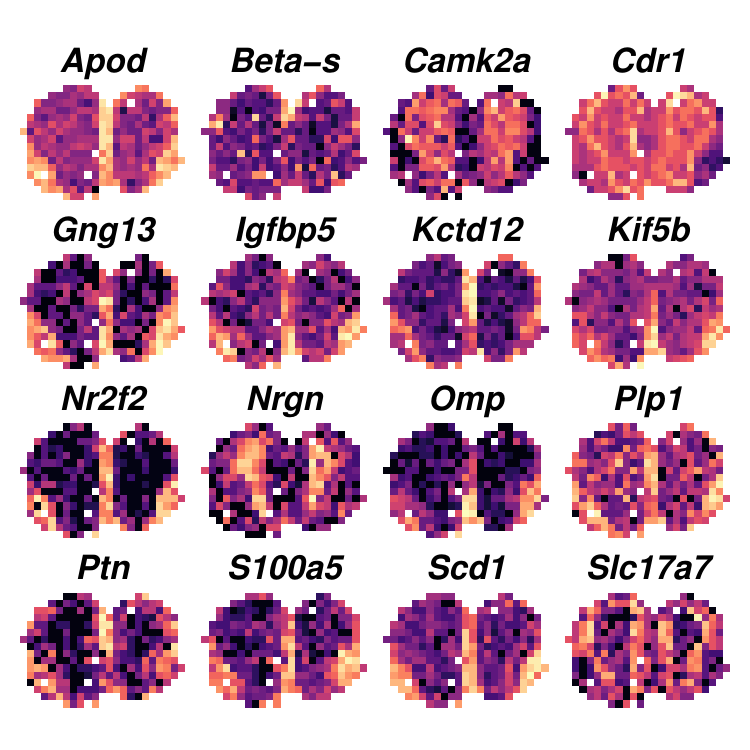}
\end{center}
\caption{MOB ST data analysis: Spatial expression patterns of the top $16$ BNPSpaced-identified DGs determined by their PPIs. \label{fig:s6}}
\end{figure}

\begin{figure}
\renewcommand*\thefigure{S\arabic{figure}}
\begin{center}
\includegraphics[width = 1\textwidth]{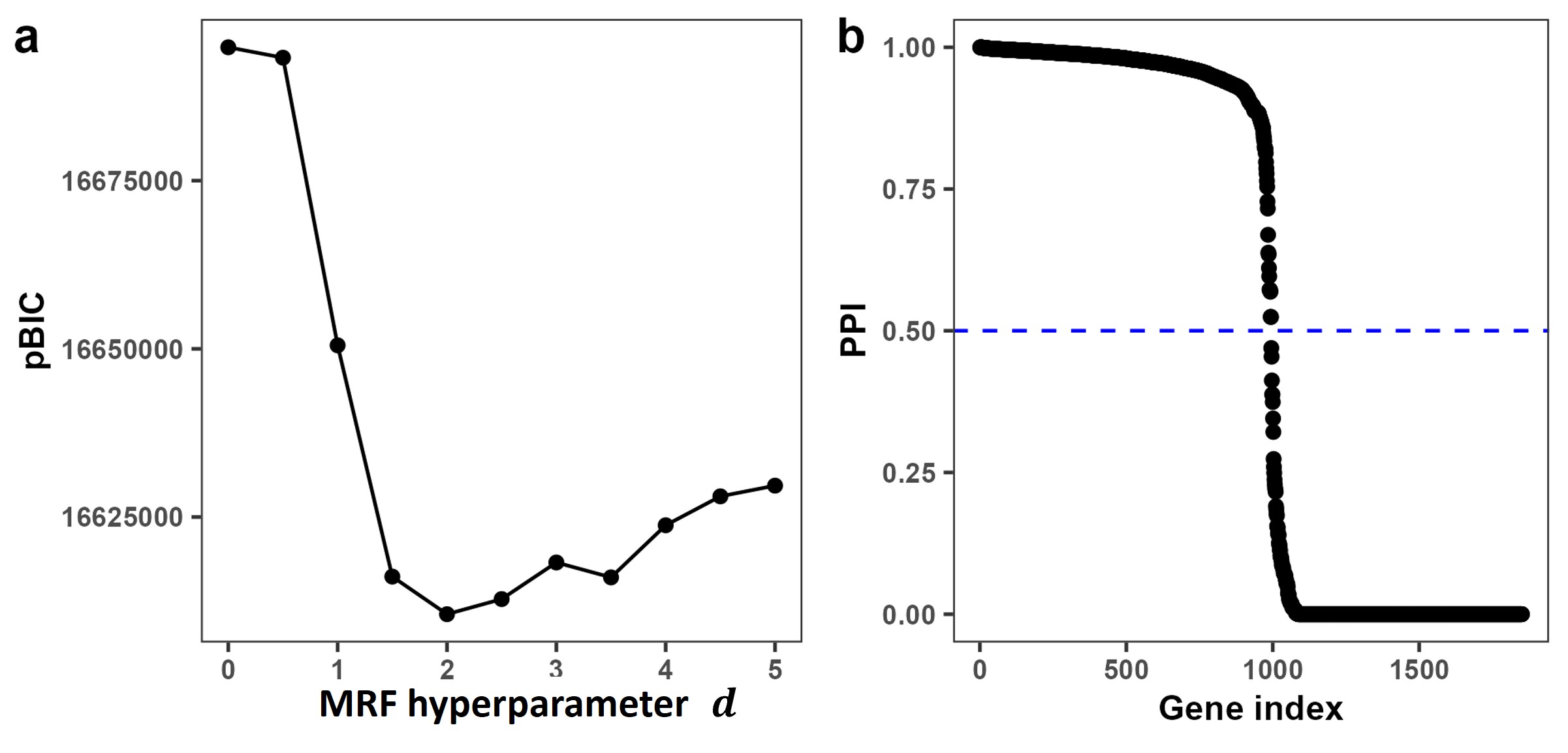}
\end{center}
\caption{Human DLPFC 10x Visium data analysis: (a) The pBIC across various values of $d$, aiding in the selection of this MRF hyperparameter; (b) The sorted PPI values for $p=1,851$ genes by BNPSpace, aiding in the selection of DGs. \label{fig:s7}}
\end{figure}

\begin{figure}
\renewcommand*\thefigure{S\arabic{figure}}
\begin{center}
\includegraphics[width = 1\textwidth]{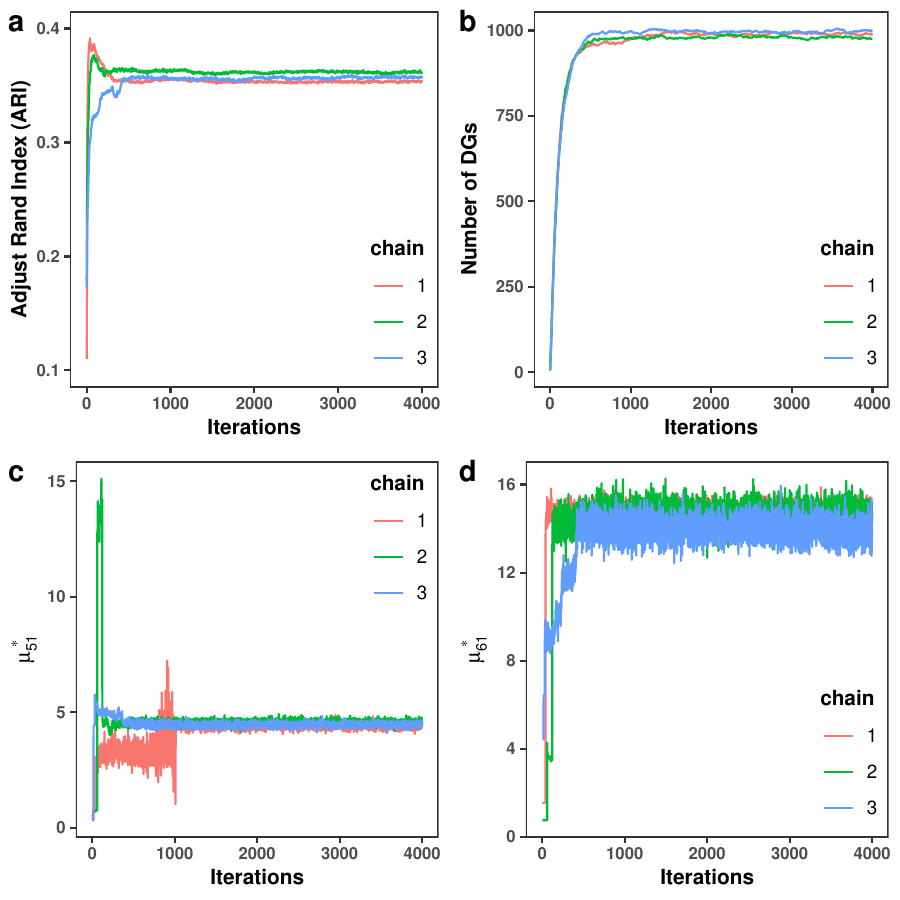}
\end{center}
\caption{Human DLPFC 10x Visium data analysis: Trace plots of (a) ARI, (b) The number of DGs $p_\gamma$, (c) the normalized expression level $\mu_{51}^*$ for gene $1$ across all spots in spatial domain $5$, and (d) the normalized expression level $\mu_{61}^*$ for gene $1$ across all spots in spatial domain $6$, of three independent MCMC chains with diverse initializations. \label{fig:s8}}
\end{figure}

\begin{figure}
\renewcommand*\thefigure{S\arabic{figure}}
\begin{center}
\includegraphics[width = 1\textwidth]{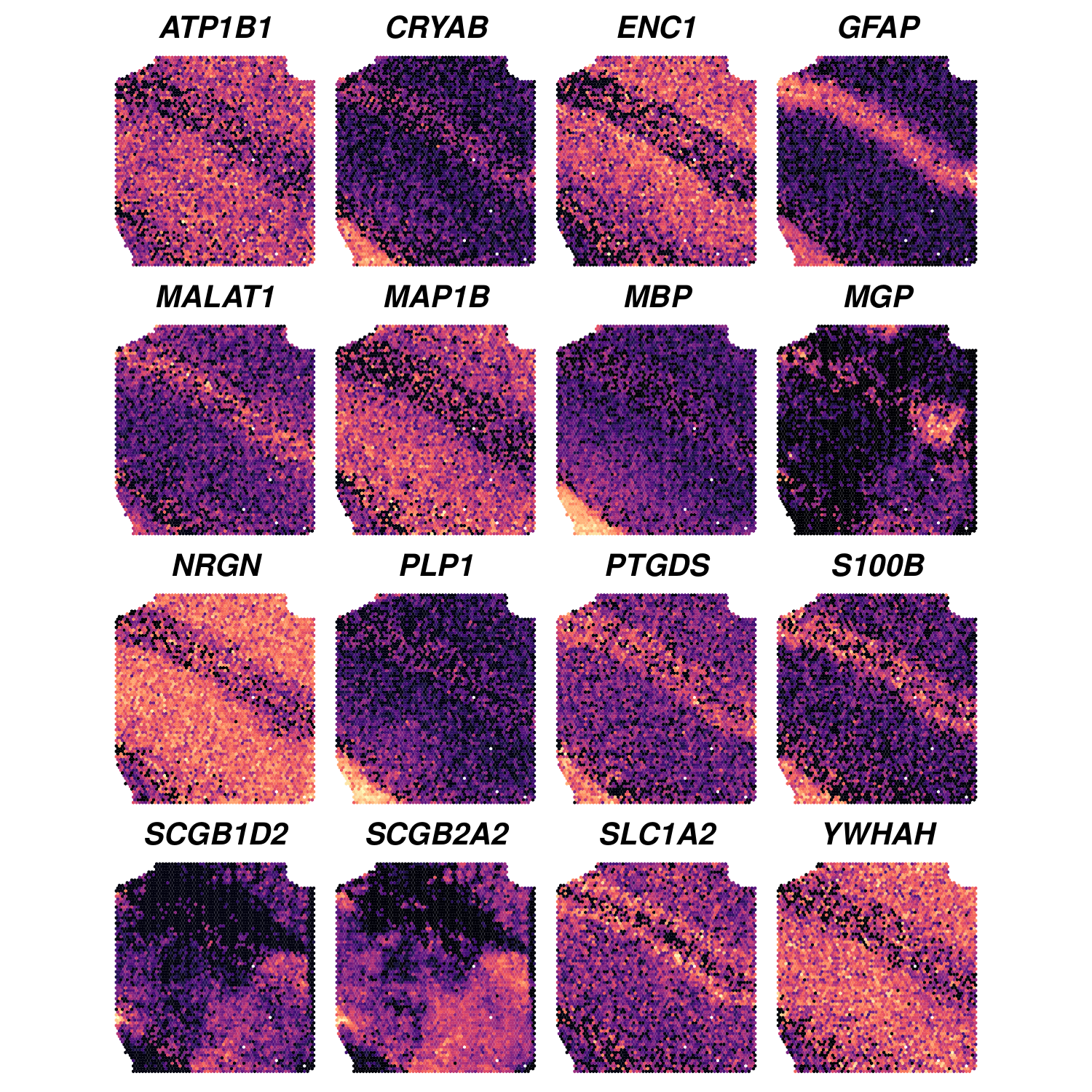}
\end{center}
\caption{Human DLPFC 10x Visium data analysis: Spatial expression patterns of the top $16$ BNPSpaced-identified DGs determined by their PPIs. \label{fig:s9}}
\end{figure}

\begin{figure}
\renewcommand*\thefigure{S\arabic{figure}}
\begin{center}
\includegraphics[width = 1\textwidth]{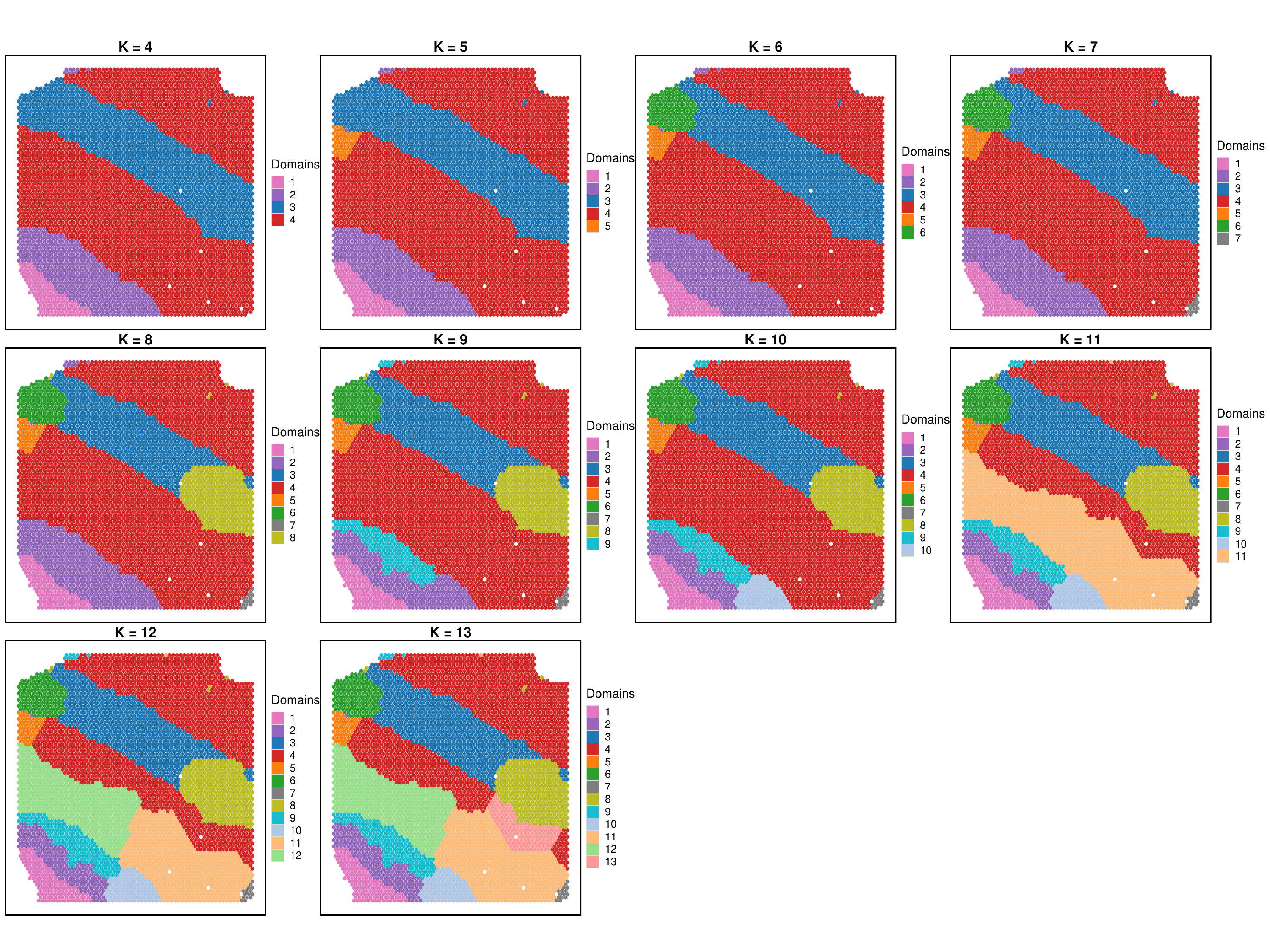}
\end{center}
\caption{Human DLPFC 10x Visium data analysis: The spatial domain patterns under different $K$ values when pruning the dendrogram based on the pairwise distances between BNPSpace-identified domains, $\text{dist}(k,k')=\sqrt{\sum_{\{j:\hat{\gamma}_j=1\}}(\hat{\mu}_{kj}^*-\hat{\mu}_{k'j}^*)^2}$. \label{fig:10}}
\end{figure}

\clearpage

\begin{table}[h]
\renewcommand*\thetable{S\arabic{table}}
\caption{Simulated data analysis: Numerical summary of averaged sensitivities, specificities, MCCs, and AUCs on for the DG indicator vector $\bm{\gamma}$, as achieved by BNPSpace across various scenarios in terms of in terms of spatial domain patterns and zero-inflation settings, using a PPI threshold at $c=0.5$. Standard deviations are indicated in parentheses.  \label{tab:s2}}
\begin{center}
    \begin{tabular}{cccccc}
\hline
\makecell[c]{\textbf{Spatial} \\ \textbf{pattern}} & \makecell[c]{\textbf{Zero-inflation} \\ \textbf{level} } & \textbf{Sensitivity} & \textbf{Specificity} & \textbf{MCC} & \textbf{AUC} \\ \hline
I            & Low          & 1.000(0.000)                 &0.986(0.004)         & 0.866(0.035) & 0.998(0.001) \\
I            & Medium          & 1.000(0.000)                 & 0.990(0.005)         & 0.902(0.043) & 0.998(0.001) \\
I           & High          & 1.000(0.000)                 & 0.992(0.003)         & 0.913(0.041) & 0.999(0.001) \\
II            & Low          & 1.000(0.000)                 & 0.998(0.002)         & 0.979(0.025) & 0.999(0.001) \\
II           & Medium          & 1.000(0.000)                 & 0.995(0.005)         & 0.951(0.049) & 0.999(0.001) \\
II           & High          & 1.000(0.000)                 & 0.992(0.009)         & 0.924(0.075) & 0.999(0.002) \\
III            & Low          & 1.000(0.000)                 & 0.999(0.001)         & 0.991(0.016) & 0.999(0.001)     \\
III            & Medium          & 1.000(0.000)                 & 0.998(0.002)         & 0.979(0.030) & 0.999(0.001) \\
III            & High          & 1.000(0.000)                 & 0.997(0.007)         & 0.966(0.057) & 0.999(0.001) \\ 
IV            & Low          & 1.000(0.000)                 & 0.996(0.001)         & 0.986(0.015) & 0.999(0.001)     \\
IV            & Medium          & 1.000(0.000)                 & 0.997(0.003)         & 0.969(0.026) & 0.999(0.001) \\
IV            & High          & 1.000(0.000)                 & 0.998(0.004)         & 0.976(0.067) & 0.999(0.001) \\
\hline
\end{tabular}
\end{center}
\end{table}

\begin{table}[h]
\renewcommand*\thetable{S\arabic{table}}
\caption{Simulated data analysis: Numerical summary of averaged sensitivities, specificities, MCCs, and AUCs on for the DG indicator vector $\bm{\gamma}$, as achieved by BNPSpace across various scenarios in terms of in terms of spatial domain patterns and zero-inflation settings, using a PPI threshold that controls a Bayesian FDR of $5\%$. Standard deviations are indicated in parentheses.  \label{tab:s3}}
\begin{center}
    \begin{tabular}{cccccc}
\hline
\makecell[c]{\textbf{Spatial} \\ \textbf{pattern}} & \makecell[c]{\textbf{Zero-inflation} \\ \textbf{level} } & \textbf{Sensitivity} & \textbf{Specificity} & \textbf{MCC} & \textbf{AUC} \\ \hline
I           & Low          & 1.000(0.000)                 & 0.987(0.004)         & 0.871(0.037) & 0.998(0.001) \\
I            & Medium          & 1.000(0.000)                 & 0.991(0.005)         & 0.909(0.050) & 0.998(0.001) \\
I            & High          & 1.000(0.000)                 & 0.993(0.005)         & 0.925(0.047) & 0.999(0.001) \\
II            & Low          & 1.000(0.000)                 & 0.989(0.097)         & 0.989(0.018) & 0.999(0.001) \\
II            & Medium          & 1.000(0.000)                 & 0.996(0.005)         & 0.961(0.050) & 0.999(0.001) \\
II            & High          & 1.000(0.000)                 & 0.992(0.011)         & 0.934(0.084) & 0.999(0.002) \\
III            & Low          & 1.000(0.000)                 & 0.999(0.001)         & 0.997(0.009) & 0.999(0.001)     \\
III            & Medium          & 1.000(0.000)                 & 0.998(0.002)         & 0.985(0.029) & 0.999(0.001) \\
III            & High          & 1.000(0.000)                 & 0.997(0.007)         & 0.975(0.057) & 0.999(0.001) \\
IV            & Low          & 1.000(0.000)                 & 0.999(0.001)         & 0.998(0.009) & 0.999(0.001)     \\
IV            & Medium          & 1.000(0.000)                 & 0.993(0.005)         & 0.978(0.019) & 0.999(0.001) \\
IV            & High          & 1.000(0.000)                 & 0.994(0.003)         & 0.985(0.047) & 0.999(0.001) \\
\hline
\end{tabular}
\end{center}
\end{table}

\begin{table}[h]
\renewcommand*\thetable{S\arabic{table}}
\caption{MOB ST data and DLPFC 10X Visium data analysis: Computation time in seconds of all methods. Computation was performed on a single thread of an E5 - 2643 v4 CPU (20 M cache, 3.40 GHz) with 256GB memory. For BNPSpace, the running time was based on $100$ iterations. \label{tab:s4}}
\begin{center}
    \begin{tabular}{ccccccc}
\hline
\textbf{Data} & \textbf{BNPSpace} & \textbf{BayesSpace} &  \textbf{SpaGCN} & \textbf{SC-MEB} & \textbf{DR-SC} & \textbf{Louvain} \\ \hline
MOB           & 36          & 8      & 8           & 10    & 3  & 0.2 \\
DLPFC          & 582        & 116   & 60          & 21     & 93  & 1.6 \\
\hline
\end{tabular}
\end{center}
\end{table}

\clearpage
\bibliographystyle{agsm}
{\small \bibliography{ref}}

\end{document}